\documentclass[%
reprint,
amsmath,amssymb,
aps,
showkeys,
]{revtex4-2}

\usepackage{graphicx}
\graphicspath{{./Figures/}}
\usepackage{dcolumn}
\usepackage{multirow}
\usepackage{bm}
\usepackage{times}
\usepackage[colorlinks = true,
linkcolor = blue,
urlcolor  = blue,
citecolor = blue,
anchorcolor = blue]{hyperref}
\usepackage{placeins}
\usepackage{upgreek}
\usepackage{listings}
\usepackage{xcolor}
\usepackage{hyperref}

\newcommand{\beginsupplement}{
	\setcounter{section}{0}
	\setcounter{table}{0}
	\setcounter{figure}{0}
	\setcounter{equation}{0}
	\renewcommand{\thetable}{S\arabic{table}}
	\renewcommand{\theHtable}{Supplemental.\thetable}
	\renewcommand{\thefigure}{S\arabic{figure}}
	\renewcommand{\theHfigure}{Supplemental.\thefigure}
	\renewcommand{\theequation}{S\arabic{equation}}
	\renewcommand{\theHequation}{Supplemental.\theequation}
	\renewcommand{\thesection}{S\arabic{section}}
	\renewcommand{\theHsection}{Supplemental.\thesection}
	
}

\definecolor{codegreen}{rgb}{0,0.6,0}
\definecolor{codegray}{rgb}{0.5,0.5,0.5}
\definecolor{codepurple}{rgb}{0.58,0,0.82}
\definecolor{backcolour}{rgb}{0.95,0.95,0.92}
\lstdefinestyle{mystyle}{
  backgroundcolor=\color{backcolour}, commentstyle=\color{codegreen},
  keywordstyle=\color{magenta},
  numberstyle=\tiny\color{codegray},
  stringstyle=\color{codepurple},
  basicstyle=\ttfamily\footnotesize,
  breakatwhitespace=false,         
  breaklines=true,                 
  captionpos=b,                    
  keepspaces=true,                 
  numbers=left,                    
  numbersep=5pt,                  
  showspaces=false,                
  showstringspaces=false,
  showtabs=false,                  
  tabsize=2
}
\begin{document}
	
\preprint{APS/123-QED}

\title{Deep Potentials for Materials Science }

\author{Tongqi Wen$^1$}
\author{Linfeng Zhang$^{2,3}$}
\author{Han Wang$^{4,5}$} \email{wang\_han@iapcm.ac.cn}
\author{Weinan E$^{3,6,7}$} 
\author{David J. Srolovitz$^{1,8}$} \email{srol@hku.hk}

\affiliation{\\
	$^1$Department of Mechanical Engineering, The University of Hong Kong, Hong Kong SAR, China\\
	$^2$DP Technology, Beijing, China\\
	$^3$AI for Science Institute, Beijing, China\\
	$^4$Laboratory of Computational Physics, Institute of Applied Physics and Computational Mathematics, Beijing, China\\
	$^5$HEDPS, CAPT, College of Engineering, Peking University, Beijing, China\\
	$^6$School of Mathematical Sciences, Peking University, Beijing, China\\
	$^7$Department of Mathematics and Program in Applied and Computational Mathematics, Princeton University, Princeton, USA\\
	$^8$International Digital Economy Academy (IDEA), Shenzhen, China
}

\date{\today}
\begin{abstract}
To fill the gap between  accurate (and expensive) \textit{ab initio} calculations and efficient atomistic simulations based on empirical interatomic potentials, a new class of descriptions of atomic interactions has  emerged and been widely applied; i.e., machine learning potentials (MLPs).  One recently developed  type of MLP is the Deep Potential (DP) method. In this review, we provide an introduction to DP methods in computational materials science. The theory underlying the  DP method is presented along with a step-by-step introduction to their development and use. We also review materials applications of DPs in a wide range of materials systems. The  DP Library provides a platform for the development of DPs and a database of extant DPs. We discuss the accuracy and efficiency of DPs compared with  \textit{ab initio} methods and empirical potentials. 
\end{abstract}


%
\maketitle

\section{\label{sec:sec1}Introduction}

Atomistic simulations are playing an increasingly important role in materials science and changing how research in this heavily-experimental field is conducted~\cite{Hafner_2000_acta}. 
All atomistic simulations in materials modelling requires the input of some form of potential energy surface (PES) to describe how atoms interact;  from which atomic forces are determined.  
The most accurate way to obtain the PES, within  the Born-Oppenheimer approximation~\cite{born1927quantentheorie},  is by solving the Schr\"odinger equation based on a quantum mechanical treatment of the electronic structure for fixed atomic nuclei coordinates~\cite{Dirac_1929_prsla}.
However, in the most-widely applied electronic structure approach, density function theory (DFT)~\cite{Kohn_1965_pr}, the scaling is normally $\propto N^3$ where $N$ is  the number of atoms. 
This scaling makes DFT  very costly in applications to large materials systems (over 1000 atoms) and for long simulation times (e.g., nanoseconds in molecular dynamics, MD). 
A widely applied solution to this efficiency problem is to develop empirical interatomic potentials~\cite{Verlet_1967_pr,Zwanzip_1954_jcp,Tersoff_1989_prb,Vink_2001_jncs,Daw_1984_prb,Baskes_1992_prb}, which describe the relationship between atom positions and system energy by assuming an analytical functional relationship, often based upon physical and chemical insights. Although large-scale, long-time atomistic simulations may readily be performed using empirical interatomic potentials, the accuracy of the atomistic simulations in describing real materials systems is often limited by the assumptions inherent to these empirical descriptions. 
In this context, one faces the dilemma that quantum mechanics methods are highly accurate but extremely inefficient for such atomistic simulations while empirical interatomic potentials are efficient, but commonly of limited accuracy.

Many approaches have been proposed to strike a balance between  accuracy and efficiency in atomistic simulations. 
One approach to overcoming the low efficiency of widely-used DFT methods was the development of the ONESTEP program~\cite{Prentice_2020_jcp} in which plane-wave  DFT calculations are performed on parallel computers, leading to computational costs that are linear in the number of atoms. 
DFT calculations have also been implemented on GPU machines~\cite{Hacene_2012_jcc,Hutchinson_2012_cpc,Jia_2013_cpc,Jia_2013_jcp} leading to accelerations by a factor of  more than 20 compared to CPU machines~\cite{Jia_2013_jcp}. 
The accuracy of empirical interatomic potentials can be improved, to some extent, by  developing analytical functional forms that include many additional parameters to account for additional physical insights. 
A typical example is the modified embedded atom method (MEAM) potential~\cite{Baskes_1992_prb} which extends the embedded atom method (EAM)~\cite{Daw_1984_prb} by considering the angular nature of electron density distributions. 
However, this tradeoff leads to a decrease in computational speed  of the MEAM potentials compared to the simpler EAM potentials; such angular potentials are also more difficult to accelerate than simpler potentials. 
Nonetheless, the progress from these two approaches are encouraging and have led to many applications. 
Nevertheless, as materials systems of interest become larger, more complex and demands on predictability more severe, obtaining a better balance between  accuracy and efficiency is both an urgent and challenging problem. 
This challenge  requires fundamentally new approaches, rather than tweaking existing techniques. 
New insights may be garnered from recent progress in other disciplines.

Machine learning (ML) is well known for its surprising successes in, for example, pattern recognition and, as a result many different ML methods were developed  in recent years~\cite{Bishop_2006_book}. 
ML potentials may be viewed as versatile descriptions of the PES parameterised using a flexible ML-based analytical form. 
The flexibility of ML potential increases the representability (accuracy) compared to  empirical interatomic potentials and the analytical form significantly improves the efficiency relative  to DFT calculations. 
In this perspective, ML potentials, fit to DFT results, have the potential for achieving DFT accuracy and empirical interatomic potential efficiency.
Developing such ML potentials is a challenge.

Since the pioneering work of Blank, et al.~\cite{Blank_1995_jcp} where ML neural network methods for describing the PES were first introduced, a variety of ML potentials have been proposed.
Behler and Parrinello (BP)~\cite{Behler_2007_prl} introduced ML neural network potential (NNP) approach in which   radial and angular symmetry functions are used as atomic environment descriptors; NNP has found applications in bulk silicon~\cite{Behler_2007_prl}, carbon~\cite{Khaliullin_2011_nm}, TiO$_2$~\cite{Artrith_2016_cms} and many other materials~\cite{Behler_2021_cr}.
BP-NNPs were categorised into four generations~\cite{Behler_2021_cr} for which there are several recent reviews~\cite{Behler_2016_jcp,Behler_2017_angew,Behler_2021_cr}. 
Sch\"utt, et al.~\cite{Schutt_2018_jcp,Schutt_2019_jctc} developed SchNet and the SchNetPack package based upon a neural network framework to model the chemical properties and PES of molecular materials. 
Many other NNPs exist and are popular for different material systems~\cite{Ghasemi_2015_prb,Hy_2018_jcp,Unke_2019_jctc,Pun_2019_nc}. 
Apart from NNP and SchNet, several other types of ML potentials were introduced;  Gaussian approximation potentials (GAP)~\cite{Bartok_2010_prl,Dragoni_2018_prm,Bartok_2018_prx,Deringer_2021_cr},  moment tensor potentials (MTP)~\cite{Shapeev_2016_mms,Podryabinkin_2017_cms,Podryabinkin_2019_prb}, spectral neighbour analysis potentials (SNAP)~\cite{Chen_2017_prm,Li_2018_prb,Deng_2019_npj}, gradient-domain machine learning (GDML)~\cite{Sauceda_2019_jcp,Chmiela_2019_cpc}, et al~\cite{Unke_2021_cr}. 
A comprehensive comparison of the major ML potentials in terms of accuracy and efficiency can be found in a recent paper~\cite{Zuo_2020_jpca}.

Deep Potentials (DP)~\cite{Zhang_2018_prl,Wang_2018_cpc} are of the NNP type that were introduced in 2018.
This approach has been used extensively for different material systems.
The underlying theory has also developed continuously pushing these potentials to increasingly favourable combinations of accuracy and efficiency.
Recently, DP has been applied to MD simulations of more than 100 million atoms with \emph{ab initio} accuracy on a state-of-the-art supercomputer~\cite{Jia_2020_sc20}. 
This is a good example of the power of integrating physical modelling, machine learning and high-performance computing. 
In this review, we focus on the application of DP in  materials science and discuss a vision for future DPs. 
The paper is organised as follows. 
In Sect.~\ref{sec:sec2}, we review the basic theory underlying the DP method (Sect.~\ref{sec:sec2a}), demonstrate the steps for developing DPs and their application for atomistic simulation (Sect.~\ref{sec:sec2b}), introduce the extant software for DP development (Sect.~\ref{sec:sec2c}) and the DP Library (Sect.~\ref{sec:sec2d}), and discuss  how to make DP more practical for atomistic simulations (Sect.~\ref{sec:sec2e}). 
Sect.~\ref{sec:sec3} lists many examples of DP applications in materials science, covering elemental bulk systems (Sect.~\ref{sec:sec3a}), multi-element bulk systems (Sect.~\ref{sec:sec3b}), aqueous systems (Sect.~\ref{sec:sec3c}), and other applications (Sect.~\ref{sec:sec3d}). 
Next, we discuss the efficiency and accuracy of DPs in practice (Sect.~\ref{sec:sec4a}) and a comparison of the computational speed of DP versus other approaches (Sect.~\ref{sec:sec4b}). 
We conclude Sect.~\ref{sec:sec5} with an  assessment of where DP  is going in the near future.        

\section{\label{sec:sec2}Deep Potential}

\subsection{\label{sec:sec2a}Theory}
Consider a system of \textit{N} atoms, where the total energy of the system is denoted by \textit{E},  the atomic coordinates by  $\mathcal{R}=\{\bm r_1,\bm r_2,...,\bm r_i,...,\bm r_N\}$ and  $\{r_{i1},r_{i2},r_{i3}\}$ are the three Cartesian components of the vector position of atom $i$, $\bm r_i$. 
The potential energy \textit{E} is  a function of all atom coordinates, i.e., $E=E(\mathcal{R})$, and can be accurately determined from first principles calculations. 
The  force on atom $i$ is the negative gradient of the potential energy with respect to its atomic coordinates ($\bm{F}_i=\{F_{i1},F_{i2},F_{i3}\}$):
\begin{eqnarray}\label{eq:f}
\bm{F_i}=-\nabla_{\bm{r_i}}E. \label{equ1}
\end{eqnarray}
Periodic boundary conditions are often applied in atomistic simulations; we denote the cell vectors by a matrix $h = \{h_{\alpha\beta}\}$, where $h_{\alpha\beta}$ is the $\beta^\text{th}$ component of the $\alpha^\text{th}$ cell vector. 
The virial tensor is then defined by $\bm \Xi = \{\Xi_{\alpha\beta}\}$ with
\begin{align}\label{eq:v}
\Xi_{\alpha\beta} = -\frac{\partial E}{\partial h_{\gamma\alpha}} h_{\gamma\beta}.
\end{align}

Training of the ML potential belongs to the category of classical supervised ML. 
We first obtain the total energy, atomic forces and virial tensors of a number of different system configurations described by the atomic coordinates and use these data as the training labels. 
Then the ML potential is trained on these labels. 
We denote a ML potential as $E^w(\mathcal{R})$ where $w$ is the set of  trainable parameters of the model. 
The force $\bm F^w(\mathcal R)$ and the virial tensor $\bm \Xi^w(\mathcal R)$ of the ML potential are derived from $E^w$ by Eq.~\eqref{eq:f} and \eqref{eq:v}, respectively. 
The derivatives can be obtained analytically if the potential $E^w(\mathcal{R})$ is differentiable with respect to $\mathcal{R}$. 
In this case, the training  of the ML potential over the training labels  can be regarded as a minimisation process of the loss function:
\begin{align}\label{eq:loss}
    \mathcal L &= \frac{1}{\vert \mathcal B\vert} 
    \sum_{k\in \mathcal B}
    \Big(
     p_e \mathcal L_e^{(k)} + p_f \mathcal L_f^{(k)} + p_v \mathcal L_v^{(k)}
    \Big), \\
    \mathcal L_e^{(k)} & = \frac 1N \big\vert E(\mathcal{R}^{(k)})-E^w(\mathcal{R}^{(k)}) \big\vert^2, \\
    \mathcal L_f^{(k)} &= \frac 1{3N} \sum_{i\alpha} \big\vert F_{i\alpha}(\mathcal{R}^{(k)})-F^w_{i\alpha}(\mathcal{R}^{(k)})  \big\vert^2,\\
    \mathcal{L}_v^{(k)} &= \frac 1{9N} \sum_{\alpha\beta} \big\vert \Xi_{\alpha\beta}(\mathcal{R}^{(k)})-\Xi_{\alpha\beta}^w(\mathcal{R}^{(k)})\big\vert^2,
\end{align}
where $\mathcal{B}$ is a mini-batch of the training data and $|\mathcal{B}|$ is the number of configurations in this batch. 
$p_e$, $p_f$, and $p_v$ are the prefactors of the energy, forces, and virials in the loss function, which are defined by user or adjusted according to the learning rate in the training process~\cite{Zhang_2018_prl}.

In the ML potential, the extensibility of the total energy is preserved upon decomposition into  atomic energies as follows:
\begin{eqnarray}
	E^w=\displaystyle\sum_{i=1}^{N}E_i^w=\displaystyle\sum_{i=1}^{N}E_i^w(\mathcal{R}_i), \label{equ7}
\end{eqnarray}
where $E_i^w$ is the  energy we associate with atom $i$. 
In the ML potential, as well as many classical interatomic potentials (such as EAM potentials), a common assumption is that the energy of atom $i$ depends only on its atomic coordinate and local environment. 
Consider $r_c$ as a pre-defined cutoff radius (the choice of $r_c$ depends on the atomic interaction characteristics of the material), the local environment of atom $i$ is the collection of the relative positions of all  neighbouring atoms whose distance to atom $i$ is smaller than $r_c$. 
This collection of near neighbours is denoted as $\mathcal{N}_{r_c}(i) = \{j| r_{ij} = \vert\bm r_{ij}\vert \leq r_c \}$.
We define the cardinality of the set $\mathcal{N}_{r_c}(i)$ as $N_i$ and use the environment matrix $\mathcal{R}_i$ with $N_i$ rows and 3 columns  to represent the local environment of atom $i$, where  row $j$ of $\mathcal{R}_i$ is the relative position of atoms $i$ and $j$:
\begin{eqnarray}
	(\mathcal{R}_i)_j=(\bm{r}_{ij}). \label{equ8}
\end{eqnarray}

The local (neighbourhood) dependence of the atomic energy  is an assumption.
There are, however,  non-local (long-range) interactions, arising mainly  from Coulombic interactions within the  electron density distribution. 
For metallic systems, the local dependence assumption is reasonable as a result of shielding effects. 
For homogeneous materials,  the long-range interactions attenuates rapidly with  increasing  atomic separation such that a sufficiently large  $r_c$ can always satisfy the local dependence assumption.
For materials where  long-range interactions dominate, these must be explicitly considered in the model construction.
Although there are research reports  which introduce long-range interaction in the model construction~\cite{Unke_2019_jctc,Artrith_2011_prb,Bereau_2015_jctc,Bereau_2018_jcp,Nebgen_2018_arxiv,Sifain_2018_jpcl,Ghasemi_2015_prb,Ko_2021_nc,Ko_2021_acr,Grisafi_2019_jcp,Grisafi_2021_cs}, there is still no widely accepted method which can handle this interaction appropriately. 
In the DP method, we focus on the most common situation, in which $E_i^w$ is well-described by the  local interaction assumption.  

The total energy of the material system is invariant under a set of symmetry operations that may include translations, rotations, and permutations:
\begin{eqnarray}
	E(\mathcal{R})=E(U\mathcal{R}), \label{equ9}
\end{eqnarray}
where $U$ is the symmetry operation on the atomic coordinates. 
For the ML potential model, the most common approach is to insure that 
\begin{eqnarray}
	E_i^w(\mathcal{R}_i)=E_i^w(U\mathcal{R}_i). \label{equ10}
\end{eqnarray} 
To achieve this energy invariance,  descriptors of the atomic environment are introduced that are invariant upon these symmetry operations  performed on the atomic coordinates:
\begin{eqnarray}
	\mathcal{D}(\mathcal{R}_i)=\mathcal{D}(U\mathcal{R}_i). \label{equ11}
\end{eqnarray}
The atomic energy can thus be written as:
\begin{eqnarray}
	E_i^w(\mathcal{R}_i)=\mathcal{F}(\mathcal{D}(\mathcal{R}_i)), \label{equ12}
\end{eqnarray}
where $\mathcal{F}$ is the fitting method employed for the  deep neural network in the DP construction.
The smoothness of the formalism is also important for calculating atomic forces and virial tensors from derivatives. 
This implies that the descriptors should be of sufficiently high resolution so as to distinguish intrinsically different local atomic environments.

Different types of ML potentials employ different descriptors.  
Some examples are the symmetry functions from Behler and Parrinello~\cite{Behler_2007_prl}, the Smooth Overlap of Atomic Positions (SOAP) from Bartok, et al.~\cite{Bartok_2018_prx}, the SchNet ``descriptors'' from Sch\"utt, et al.~\cite{Schutt_2018_jcp,Schutt_2019_jctc}, and the moment tensor from Shapeev~\cite{Shapeev_2016_mms}.      
In the following, we  introduce the deep neural network (DNN) and then focus on the construction and physical meanings of the descriptors employed in DPs.   

\subsubsection{\label{sec:sec2a1}Deep neural network}%
Due to the robust fitting ability of deep neural networks (DNN) for high-dimensional and nonlinear properties, the DP uses DNNs both as a fitting method for $\mathcal{F}$ and also to construct descriptors of local atomic environments (as described below, Section~\ref{sec:sec2a3}) 

The DNN is  used as the fitting net $\mathcal{F}$ that maps the descriptors $\mathcal{D}_i$ to local atomic energies:
\begin{eqnarray}
	\mathcal{F}(\mathcal{D}_i)=\mathcal{L}_O \circ \mathcal{L}_P \circ \cdots \circ \mathcal{L}_k \circ \cdots \circ \mathcal{L}_1 (\mathcal{D}_i), \label{equ13}
\end{eqnarray}
where $\mathcal{L}_k$ is the mapping from layer $k-1$ to $k$ ($P$ hidden layers in total) in the neural network.
Each mapping is composed of a linear and a non-linear transformation as follows:
\begin{eqnarray}
	\bm{d}_{k}=\mathcal{L}_k(\bm{d}_{k-1})=\phi(\mathcal{W}_k \cdot \bm{d}_{k-1}+\bm{b}_k), \label{equ14}
\end{eqnarray}
where $\bm{d}_{k} \in \mathbb{R}^{n_k}$ represents the state of the  neurons in layer $k$, and $n_k$ is the number of neurons. 
The weight matrix $\mathcal{W}_k \in \mathbb{R}^{n_k \times n_{k-1}}$ and bias vector $\bm{b}_{k} \in \mathbb{R}^{n_k}$ are trainable parameters in the neuron network. 
$\phi$ is a non-linear activation function, e.g., a hyperbolic tangent.
$\mathcal{L}_O$ is the output mapping from the last hidden layer to the output layer and is usually a linear function with trainable parameters.
From Eq.~\eqref{equ14}, we see that the smoothness of thee DNN is determined by the activation function. 
The addition of ``skip'' connections  can improve the accuracy~\cite{He_2016_CVPR}:
\begin{eqnarray}
	\bm{d}_{k}=\bm{d}_{k-1}+\phi(\mathcal{W}_k \cdot \bm{d}_{k-1}+\bm{b}_k). \label{equ15}
\end{eqnarray}
Quantification of the representability of a DNN is an active area of research. 
Barron~\cite{Barron_1993_itit,Barron_1994_ml} proved that a neural network with only 1 hidden layer ($P$=1) and an arbitrary number of neurons can approximate a class of functions with arbitrary precision. 
Many researchers have subsequently investigated the approximation ability of DNNs with $P > 1$ ~\cite{Liang_2017_arxiv,Feldman_2016_aclt,Yarotsky_2017_nn,Lu_2021_jma} and explained why the DNN is more successful (and widely used)  than  wide neural networks (many neurons in one hidden layer). 
E, et al.~\cite{E_2019_cms,E_2021_ca} explained and demonstrated why DNNs are suitable for  high dimensional problems.

\subsubsection{\label{sec:sec2a2}Descriptors for the non-smooth DP}
There are two classes of descriptors for DPs, namely  non-smooth~\cite{Zhang_2018_prl,Wang_2018_cpc} and smooth mappings of the atomic coordinates~\cite{Zhang_2018_nips}. 
The concept behind the non-smooth descriptor is to set up a local coordinate frame for every atom and its neighbours inside the cutoff distance $r_c$ and then sort neighbour atoms according to the distance to the centre atom. 
This helps preserve the translational, rotational, and permutational symmetries of the atomic environment. 
Construction of the local frame is as follows. 
First, two atoms are chosen from the neighbours of atom $i$: $a(i) \in \mathcal{N}_{r_c}(i)$, $b(i) \in \mathcal{N}_{r_c}(i)$ such that atoms $i$, $a(i)$, and $b(i)$ are not colinear. 
We define the rotational matrix as follows and the elements in each column are the basis vectors in the local coordinate system:
\begin{eqnarray}
	\mathcal{R}(\bm{r}_{ia(i)},\bm{r}_{ib(i)})= 
	\begin{pmatrix}
		\bm{e}(\bm{r}_{ia(i)}) \\
		\\
		\bm{e}[\bm{r}_{ib(i)}-\frac{\bm{r}_{ia(i)} \cdot \bm{r}_{ib(i)}}{\bm{r}_{ia(i)} \cdot \bm{r}_{ia(i)}}\bm{r}_{ia(i)}]\\
		\\
		\bm{e}(\bm{r}_{ia(i)} \times \bm{r}_{ib(i)})
	\end{pmatrix}^T \label{equ16}
\end{eqnarray}
where $\bm{e}(\bm{r})=\bm{r}/|\bm{r}|$ is the normalised vector of $\bm{r}$.
Then the local coordinates $\bm r'_{ij}=(x'_{ij},y'_{ij},z'_{ij})$ can be transformed from the global coordinates $\bm r_{ij}=(x_{ij},y_{ij},z_{ij})$ according to
\begin{eqnarray}
	(x'_{ij},y'_{ij},z'_{ij})=(x_{ij},y_{ij},z_{ij}) \cdot \mathcal{R}(\bm{r}_{ia(i)},\bm{r}_{ib(i)}). \label{equ17}
\end{eqnarray}
The local coordinates can be used directly to construct the descriptor by using radial and/or angular information:
\begin{eqnarray}
	\{D_{ij}\} = \begin{cases}
		\Big\{\frac{1}{r_{ij}},\frac{x'_{ij}}{r_{ij}},\frac{y'_{ij}}{r_{ij}},\frac{z'_{ij}}{r_{ij}}\Big\}^{\text{sort}}_{j \in \mathcal{N}_{r_c}(i)}, &\text{full info.}\\
		\\
		\Big\{\frac{1}{r_{ij}}\Big\}^{\text{sort}}_{j \in \mathcal{N}_{r_c}(i)}, &\text{radial-only}
	\end{cases} \label{equ18}
\end{eqnarray}
where the superscript ``sort" implies sorting the atoms according to their inverse distance to atom $i$ (i.e., $1/r_{ij}$). 
When a neighbour atom is far away from the centre atom, the radial-only information can be considered rather than both angular and radial information. 

\begin{figure}[]
	\includegraphics[width=0.4\textwidth]{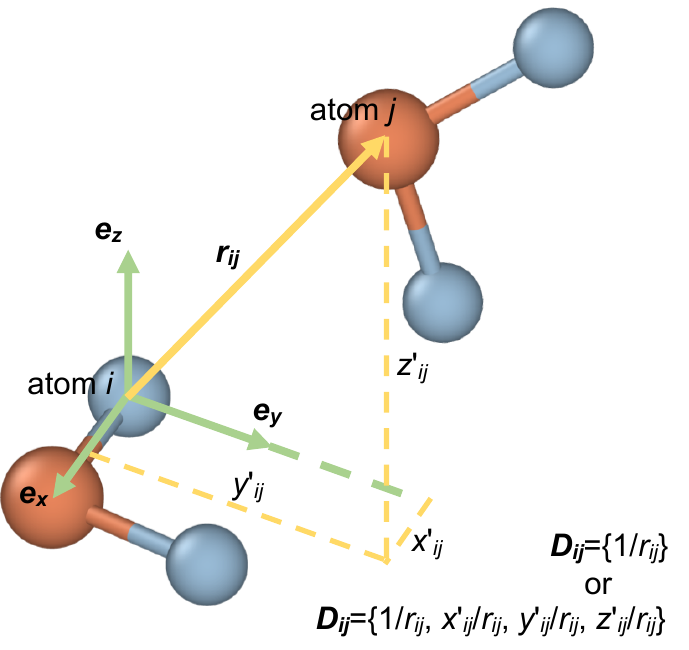}
	\caption{\label{fig:fig1}Schematic  of the non-smooth DP descriptor  for a water molecule. The red and blue spheres denote oxygen and hydrogen atoms, respectively.}
\end{figure}

A schematic of the descriptor for the non-smooth DP is shown for a water molecule in Fig.~\ref{fig:fig1}. 
The red and blue spheres denote oxygen  and  hydrogen atoms, 
($\bm{e}_x$, $\bm{e}_y$, $\bm{e}_z$) is the local frame of atom $i$ and its neighbour atom $j$, and
$(x'_{ij},y'_{ij},z'_{ij})$ are the Cartesian components of $\bm{r_{ij}}$ in this local frame. 
The descriptor $\mathcal{D}_{ij}$ may contain  only  radial (distance) information and/or angles. 
In the water molecule example,  neighbours of atom $i$ are first sorted according to their chemical species (oxygen  first, then hydrogen). 
Then within the same species, the inverse distances to atom $i$ are used to sort the atoms. 
Finally, $\mathcal{D}_{ij}$ is applied to the sorted input data for atom $i$.

The advantage of  non-smooth descriptor is that all  neighbour information is preserved. 
However, due to uncertainty in the choice of  neighbour atoms $a(i)$ and $b(i)$, the descriptor is non-smooth. 
In practice, $a(i)$ is  picked as the nearest neighbour and $b(i)$ as the second nearest neighbour. 
Continuous change in atom positions can thus result in a discontinuous change of the atom number, the local frame, and  the local coordinates. 
In addition, the sorting operation for the other neighbours  introduces additional discontinuities in the descriptor and its  derivatives  (see Eq.~\eqref{equ18}).

\subsubsection{\label{sec:sec2a3}Descriptors for the smooth DP}
\begin{figure}[]
	\includegraphics[width=0.48\textwidth]{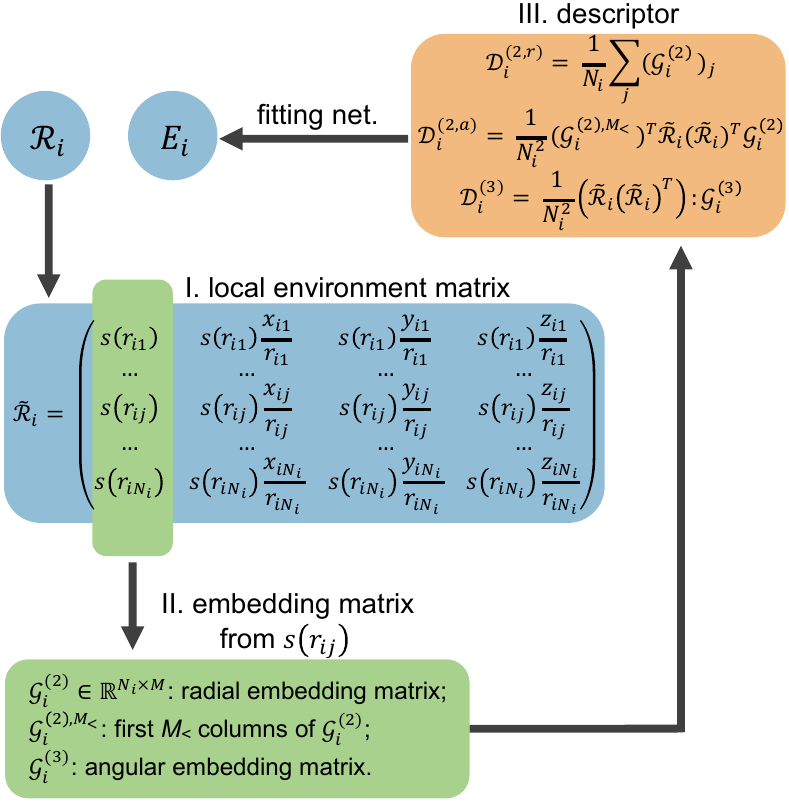}
	\caption{\label{fig:fig2}Workflow for constructing the smooth DP descriptors.}
\end{figure}

The workflow for the DP-Smooth Edition (DP-SE) model~\cite{Zhang_2018_nips} descriptor is shown in Fig.~\ref{fig:fig2}.
There are three major steps to construct the sub-network for the atomic energy using this descriptor. 
First, the environment matrix for atom $i$, Eq.~\eqref{equ8} is reformed as:
\begin{eqnarray}
	(\mathcal{R}_i)_j= s(r_{ij}) \times (\frac{x_{ij}}{r_{ij}},\frac{y_{ij}}{r_{ij}},\frac{z_{ij}}{r_{ij}}), \label{equ19}
\end{eqnarray}
where $s(r_{ij})$ is a continuous and differentiable function
\begin{eqnarray}
	s(r_{ij}) = \begin{cases}
		\frac{1}{r_{ij}}, &r_{ij} < r_{cs}\\
		\frac{1}{r_{ij}} f_c(r_{ij}), &r_{cs} < r_{ij} < r_{c}\\
		0, &r_{ij} > r_{c}\\
	\end{cases} \label{equ20}
\end{eqnarray} 
and where $f_c(r_{ij})$ is a smooth switching function decaying from 1 at $r_{cs}$ to 0 at $r_c$. 
DP uses a  $5^\text{th}$ order polynomial for this purpose: 
\begin{align}
    f_c (r) = u^3 (-6u^2 + 15 u - 10 ) + 1, \ u = \frac{r-r_{cs}}{r_c - r_{cs}}. \label{equ21} 
\end{align}
Using this $f_c$ yields a second order differentiable $s(r)$ with continuous derivatives at $r_{cs}$ and $r_c$.
The augmented matrix $\tilde{\mathcal{R}}_i \in \mathbb{R}^{N_i \times 4}$ ($N_i$ neighbours  of atom $i$) is then constructed from $\mathcal{R}_i$ by adding a column and each row is defined as:
\begin{eqnarray}
	(\tilde{\mathcal{R}}_i)_j= s(r_{ij}) \times (1,\frac{x_{ij}}{r_{ij}},\frac{y_{ij}}{r_{ij}},\frac{z_{ij}}{r_{ij}}). \label{equ22}
\end{eqnarray}

Next, the two-body embedding matrix $\mathcal{G}^{(2)}_i \in \mathbb{R}^{N_i \times M}$ is constructed from the first column in $\tilde{\mathcal R}_i$ and each row is defined as:
\begin{eqnarray}
	(\mathcal{G}_i^{(2)})_j=(G_1^{(2)}(s(r_{ij}),Z_j),...,G_M^{(2)}(s(r_{ij}),Z_j)), \label{equ23}
\end{eqnarray}
where the vector $(G_1^{(2)},...,G_M^{(2)})$ is a DNN mapping from a scalar input $s(r_{ij})$ and $Z_j$ is the input type of neighbour atom $j$.
The three-body embedding tensor is constructed as
{\footnotesize
\begin{eqnarray}
	(\mathcal{G}_i^{(3)})_{j,k}=(G_1^{(3)}((\theta_{i})_{jk},Z_j,Z_k),...,G_M^{(3)}((\theta_{i})_{jk},Z_j,Z_k)), \label{equ24}
\end{eqnarray}}
where the angular information with atom $i$ at the centre and atoms $j$ and $k$ as neighbours is the inner product of rows $j$ and $k$  of $\mathcal{R}_i$:
\begin{eqnarray}
	(\theta_{i})_{jk}=(\mathcal{R}_i)_j \cdot {(\mathcal{R}_i)_k}, \label{equ25}
\end{eqnarray}
and the vector $(G_1^{(3)},...,G_M^{(3)})$ is represented by a DNN. 

Finally, from Eq.~\eqref{equ23} and \eqref{equ24}, smooth descriptors can be constructed from the two-body and three-body embedding matrix and tensor:
\begin{align} \label{equ26}
    &\mathcal{D}_i^{(2,r)}  =\frac{1}{N_i}\sum_{j}(\mathcal{G}_i^{(2)})_j  \\\label{equ27}
    &\mathcal{D}_i^{(2,a)}  =\frac{1}{N^2_i}(\mathcal{G}_i^{(2),M_<})^T\tilde{\mathcal R}_i(\tilde{\mathcal R}_i)^T\mathcal{G}_i^{(2)} \\\label{equ28}
    &\mathcal{D}_i^{(3)}  =\frac{1}{N^2_i}(\tilde{\mathcal R}_i(\tilde{\mathcal R}_i)^T):\mathcal{G}_i^{(3)},
\end{align}
where $\mathcal{G}_i^{(2),M_<}$ are the first $M_<$ columns of $\mathcal{G}_i^{(2)}$. 
The two-body embedding $\mathcal{D}^{(2,r)}$ only depends on the radial distance between neighbouring atoms. 
While the two-body embedding $\mathcal{D}^{(2,a)}$ depends on the coordinates of the neighbour atoms,  the embedding term (Eq.~\eqref{equ23}) only relies on the atom separation. 
Three-body embedding $\mathcal{D}_i^{(3)}$ considers the angle between neighbour atoms in the embedding term (Eq.~\eqref{equ24}). 
In this case, from the point of view of descriptor accuracy and resolution, $\mathcal{D}_i^{(3)} > \mathcal{D}_i^{(2,a)} > \mathcal{D}_i^{(2,r)}$.
Training the descriptor with higher resolution is  more difficult. 
In practice, one, two or all of the descriptors above can be used in a hybrid format. 
For example, one may define the descriptor as $\mathcal D = (\mathcal{D}_i^{(3)}, \mathcal{D}_i^{(2,a)})$, where the $\mathcal{D}_i^{(3)}$ with a smaller cut-off describes near-neighbour configurations, while $\mathcal{D}_i^{(2,a)}$ with a larger cut-off  describes the environment further away.

\subsection{\label{sec:sec2b}Developing and Applying  DPs} 
After illustrating the descriptors and DNN fitting in the DP method,  we now outline the main steps we apply to develop a DP: preparing training datasets, the training process, and model validation. We also explain how to apply DPs in atomistic simulations.

\subsubsection{\label{sec:sec2b1}Preparing training datasets}
The preparation of the training data for a DP  has two parts: (1) providing atomic configurations, (i.e.,~the coordinates of atoms and the cell shape tensor) and 
(2) labelling (i.e.,~calculating the energy, atomic forces and virial tensor for the configuration). 
Labelling is done through a DFT calculation; e.g., well-established DFT packages such as  the Vienna Ab initio Simulation Package (VASP)~\cite{Kresse_1996_cms,Kresse_1996_prb}, Quantum ESPRESSO (QE)~\cite{Giannozzi_2009_jpcm}, and Atomic-orbital Based Ab-initio Computation at UStc (ABACUS)~\cite{Chen_2010_jpcm}. 
Since  DPs are trained using a DFT training set, the accuracy of a DP will never exceed that of its training data.
The errors in  DFT calculations have two main sources.
The first is the error introduced by the approximate form of the exchange-correlation functional. 
This type of error may be reduced by moving up the Jacob's ladder~\cite{Perdew_2001_acp}  of increasingly accurate exchange correlation functionals (usually at higher computational cost). 
It is also possible to use the high order post-Hartree Fock methods like the M\o ller-Plesset perturbation~\cite{Moller_1934_pr}, coupled cluster~\cite{Cizek_1966_jcp}, and conﬁguration interaction~\cite{Fano_1961_pr} for labelling.
The other source of error in DFT calculations is  numerical, i.e., error introduced by the numerical discretisation of wave functions in real and $k$ spaces and convergence. 
This type of error can be   systematically controlled  with the use of more complete basis sets (increasing the energy cutoff in  plane-wave approximations), reducing  $k$-space  mesh spacing  and using stricter stop criteria for self-consistent field (SCF) iterations.
Generally speaking, increasing label  quality implies larger computational demands. 
One usually seeks for a balance between  quality and  cost.

The training dataset is another critical issue for generating a DP;  
here, the two main issues are (a) completeness and (b) compactness.
By completeness, we mean the training datasets need to sample the relevant configuration space as completely  as possible. 
Increasing the diversity in the members of the training dataset helps increase the transferability of the DP. 
By compactness, we mean the training data should be the minimal subset of the sampled configurations from which a model with uniform accuracy on the sampled configurations is trained. 
This is important for minimising  DFT computational time.
Different approaches have been used to sample the configuration space including MD simulations,  genetic algorithms~\cite{Deaven_1995_prl,Glass_2006_cpc}, enhanced sampling methods~\cite{Laio_2002_pnas}, active learning~\cite{Cohn_1994_ml} and  concurrent learning schemes~\cite{Zhang_2019_prm,Zhang_2020_cpc}.  
Among these, the concurrent learning scheme we refer to as the deep potential generator (DP-GEN) is found to be particularly effective at generating training datasets that satisfy both the completeness and compactness conditions (see Sect.~\ref{sec:sec2c3}).

\subsubsection{\label{sec:sec2b2}Training process}
We employ the DeePMD-kit package~\cite{Wang_2018_cpc} to train a DP. 
A python package, dpdata~\footnote{\url{https://github.com/deepmodeling/dpdata}} is  used together  with the DeePMD-kit to transform the DFT labels  from DFT software to the data format accepted by the DeePMD-kit package.
DeePMD-kit and dpdata package details are discussed in Sect.~\ref{sec:sec2c}.

There are several important issues in the training process. 
First, the DP  should be  trained such as to avoid under-fitting and over-fitting.
Under-fitting implies that the DP performs poorly for both the training and validating datasets. 
Increasing the number of fitting parameters (wider or deeper deep neural networks) or adjusting the machine learning algorithms can help avoid under-fitting.
Under-fitting is  easily detected in the training process. 
On the other hand,  over-fitting  suggests that a DP  is very good at reproducing the training data, but poor at prediction. 
This may be solved by increasing the  training dataset size, decreasing the number of fitting parameters, or using the force and virial as labels (see below).
Over-fitting is more difficult to identify; hence, comprehensive testing is necessary. 

\begin{figure}
    \centering
    \includegraphics[width=0.45\textwidth]{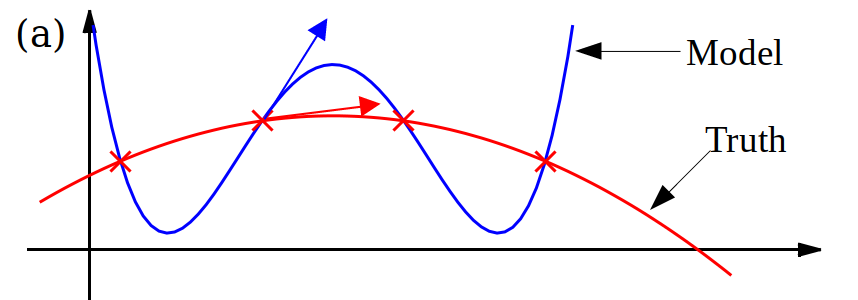}\\
    \includegraphics[width=0.45\textwidth]{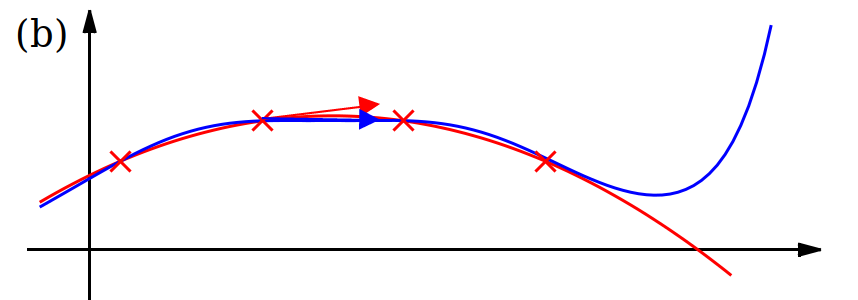}
    \caption{Schematic depiction of an over-fit ML model (a) and a properly fit ML model (b). 
    The ground truth and ML model are denoted by the red and blue lines. The training data are denoted by  red $\times$. 
    The  over-fit values are close to the ground truth for  the training data, while the gradients deviate from the ground truth (red and blue arrows).
    Both the values and gradients of the properly fit model are close to the ground truth for  the training data. 
    }
    \label{fig:fig3}
\end{figure}

Second, while training may involve the energy, force and virial labels (normally from DFT calculations),  not all of these are necessary.
However, it is strongly recommended to use force labels because (a) force has more information:  $3N$ vs.~1 compared to the energy label and (b) training with gradient information helps avoid over-fitting.
The schematic plot in Fig.~\ref{fig:fig3} demonstrates that the gradients of an over-fit model deviate from the ground truth, thus training with forces (gradients of energy with respect to coordinates) and virials (gradients of energy with respect to the cell tensor) as labels helps avoid over-fitting.

Third, there are many tunable hyper-parameters in the training process in the DeePMD-kit, including neural network size, learning rate, prefactors in the loss functions (Eq.~\eqref{eq:loss}), \ldots.  
In practice, we observe that the quality of the DP is not very sensitive to these  hyper-parameters and the default settings in the DeePMD-kit routinely provide reasonable accuracy.
In some cases, high accuracy for energy is required and ``fine tuning"  with larger energy prefactors help. 
A more detailed  discussion of the hyper-parameters in DeePMD-kit is in Sect.~\ref{sec:sec2c2}.

\subsubsection{\label{sec:sec2b3}Model validation}
After the training process is complete, it is advisable to validate the obtained DP to decide whether additional training datasets are required prior to atomistic simulation applications. 
There are  two main approaches to test the DP performance. 
(i) We can construct a small set of datasets which are unrelated to the training datasets. 
The DP  is used to predict the total energy, atomic forces, and virial tensors of the constructed datasets and the results are compared  with  DFT  results for the same atomic configuration. 
Experience shows that  the  root mean square error (RMSE) for energy and forces for a good DP  should  be smaller than $\sim$10 meV/atom and $\sim$100 meV/\AA. 
It is not uncommon to achieve  $\sim$1 meV/atom and $<$50 meV/{\AA} errors for energy and forces.
(ii) The DP can be applied to calculate properties of interest to be  compared with DFT. 
For example, elastic constants are important properties for structural applications that can be obtained both from DFT and DP MD. 
Application of DPs in atomistic simulations are discussed below and the Autotest package in DP-GEN conveniently calculates several different properties (see Sect.~\ref{sec:sec2c3}).

\subsubsection{\label{sec:sec2b4}Model inference}
Model inference (a  commonly used term in the ML community), is the process of providing live data to the ML model to obtain an output. 
In the context of ML models for the potential energy surface, the inference means taking the configuration (atom coordinates and cell tensor) as input to calculate the energy, force and virial tensor. 
With the interfaces provided by the DeePMD-kit package, one can easily make inferences about the DP in a Python or C++ programming environment.
This makes it possible to use DPs  in various molecular simulation tasks such as MD, Monte Carlo, geometric optimisation, \ldots by interfacing DeePMD-kit with molecular simulation packages,  such as the Large-scale Atomic/Molecular Massively Parallel Simulator (LAMMPS)~\cite{Plimpton_1995_jcp}, the Atomic Simulation Environment (ASE)~\cite{Larsen_2017_jpcm},  i-PI~\cite{Ceriotti_2014_cpc}, and GROMACS~\cite{Spoel_2005_jcc}. 

\subsection{\label{sec:sec2c}Software for DP development}      
There are three main packages for DP  development: (i) dpdata for converting the output of DFT software to the data format accepted by DeePMD-kit; (ii) DeePMD-kit for DP  training and inference; (iii) DP-GEN for efficient sampling and labelling of the training data. 
The DP-GEN package integrates with DeePMD-kit and DFT calculation packages to automatically generate and test DPs.

\subsubsection{\label{sec:sec2c1}dpdata}
{dpdata} is a python package which converts and manipulates training data in different DFT package formats  to the compressed format used by DeePMD-kit. 
A typical {dpdata} workflow is as follows. 
(i) Load data from data files. Data files can be written in one of the following package formats: VASP~\cite{Kresse_1996_cms,Kresse_1996_prb}, LAMMPS~\cite{Plimpton_1995_jcp}, Gaussian~\cite{Frisch_2016_Gaussian}, SIESTA~\cite{Soler_2002_jpcm}, CP2K~\cite{Kuhne_2020_jcp}, QE~\cite{Giannozzi_2009_jpcm}, FHI-aims~\cite{Blum_2009_cpc}, QUIP~\cite{QULP}, PWmat~\cite{Jia_2013_cpc,Jia_2013_jcp}, AMBER~\cite{Case_2005_jcc}, GROMACS~\cite{Spoel_2005_jcc}, and ABACUS~\cite{Chen_2010_jpcm}.
(ii) Data may be manipulated through operations including replication of atom configurations in a supercell, perturbation of the cell vector and atom positions, and replacement of a number of one type of atoms with others.
(iii) Output data in a format of one of the aforementioned software. 
Here we give a simple example of transforming the OUTCAR file from VASP into a training dataset that DeePMD-kit can read:
\begin{lstlisting}[
language=Python
]
import dpdata
d_outcar=dpdata.LabeledSystem("OUTCAR")
d_outcar.to("deepmd/npy", "dpmd_raw")
\end{lstlisting}
After execution of these commands, a directory named \lstinline{dpmd_raw} is created that stores the \lstinline{numpy} compressed data appropriate for DeePMD-kit use. 
The dpdata package source code and manual is available at~\footnotemark[1].

\subsubsection{\label{sec:sec2c2}DeePMD-kit}

The DeePMD-kit package~\cite{Wang_2018_cpc} (first made publicly in November 2018) is available  on GitHub~\footnote{\url{https://github.com/deepmodeling/deepmd-kit}} under the GNU Lesser General Public License (LGPL).  
DeePMD-kit has continued to evolve since first becoming available. 
The DeePMD-kit package interfaces with TensorFlow~\cite{tensorflow2015-whitepaper} to make the training and inference codes more  efficient and automated. 
The training is integrated and made available via a command line interface.
Model inference is provided through C++ and Python interfaces, which accept atom positions and the cell tensor and return the energy, force and virial for this configuration.
These interfaces may be used by MD and molecular simulation packages written in C, C++ or Python, as discussed in Sect.~\ref{sec:sec2b4}.
The DeePMD-kit supports GPU accelerated training and inference. 
When interfaced with the LAMMPS MD  package, parallel and distributed computations  accelerated by GPUs are available. 
(On Summit, one of the most powerful supercomputers in the world, DeePMD-kit has pushed the limit of MD with \emph{ab initio} accuracy to 100 million atoms and achieved a peak performance of 91 PetaFLOPS in double precision (45.5\% of the theoretical peak)~\cite{Jia_2020_sc20}.)
The DeePMD-kit package is now well-developed, been installed over 30,000 times (GitHub) and received over 650 stars GitHub stars.

DeePMD-kit can now be easily installed via off-line packages and package managers such as Conda and Docker. 
A detailed introduction can be found at~\footnote{\url{https://docs.deepmodeling.org/projects/deepmd/en/master/install/index.html}}. The brief release history and key milestones of DeePMD-kit can be found in the Supplemental Materials.

\paragraph{Training and testing of DPs.} 


After installation of DeePMD-kit package, we  follow the path in Sect.~\ref{sec:sec2b} to develop a DP.
First, dpdata is used to generate the training datasets to be read by DeePMD-kit package (see Sect.~\ref{sec:sec2c1}). 
Then, DP training is  started with the following command:
\begin{lstlisting}[
language=bash]
dp train input.json
\end{lstlisting}
where \lstinline{input.json} is the input script controlling the training process.
If the training program stops, it  can be restarted with 
\begin{lstlisting}[
language=bash]
dp train --restart model.ckpt input.json
\end{lstlisting}
where \lstinline{model.ckpt} is the checkpoint file storing the existing model and  training status. 
A detailed explanation of the output files and training (validation) errors can be accessed at~\footnote{\url{https://docs.deepmodeling.org/projects/deepmd/en/master/train/training.html}}.

After the training process, the architecture and the parameters of the DP can be abstracted from the checkpoint file and saved in the DP  file in the Google's protobuf format
\begin{lstlisting}[
language=bash]
dp freeze -o graph.pb
\end{lstlisting}
where \lstinline{graph.pb} is the DP file. 
The DP file can be used for testing and model inference via the Python and C++ interfaces. 
For example, the DP  can be used to give energy, forces, and virial errors on a designated datasets:
\begin{lstlisting}[
language=bash]
dp test -m graph.pb -s /path/to/system -n 30
\end{lstlisting}
where the tested model follows \lstinline{-m} tag, \lstinline{-s} gives the path to the tested system, and \lstinline{-n} is the number of frames in the tested system.

\paragraph{Fine-tuning a DP.}

The adjustment of hyper-parameters in machine learning is never a trivial issue.
Fortunately, the efficiency of training a DP  is not very sensitive to  hyper-parameters and the default settings in DeePMD-kit usually yield a DP of reasonable accuracy. 
Detailed documentation of all  available training parameters can be accessed at \footnote{ \url{https://docs.deepmodeling.org/projects/deepmd/en/master/train/training-advanced.html}}.
On occasion, the  hyper-parameters settings may not be satisfactory; e.g., if one wants a DP  with higher energy and virial accuracy without a large penalty in  force accuracy,  fine-tuning of the hyper-parameters may be necessary.

To fine-tune a DP, one may adjust the prefactors of the energy ($p_e$), forces ($p_f$), and virials  ($p_v$) in the loss function in Eq.~\eqref{eq:loss} and the learning rate.
DeePMD-kit implements an exponentially decaying learning rate and dynamically adjusts prefactors:
\begin{align}\label{eq:lr}
& r_l(t) = r_l^0 e^{t / t_d} \\\label{eq:prefactor}
&    p(t)=p^{\rm limit}[1-\frac{r_l(t)}{r_l^0}]+p^{\rm start}[\frac{r_l(t)}{r_l^0}],
\end{align}
where $t$ and $t_d$ denote the training step and the typical time-scale of the learning rate decay, respectively. 
In Eq.~\eqref{eq:prefactor}, $r_l(t)$ and $r_l^0$ are the learning rate at training step $t$ and the learning rate at the beginning, respectively. 
DeePMD-kit lets the user set the start learning rate $r_l^0$, number of training steps $T_{\textrm{train}}$ and the expected learning rate at the end of training $r_l(T_{\textrm{train}})$, then determines the parameter $t_d$ automatically.
The prefactors gradually change from $p^{\rm start}$ to $p^{\rm limit}$ in the training process. 
The default settings of the start and stop learning rates are $10^{-3}$ and $10^{-8}$, respectively, and those of the prefactors are 
$p_e^{\rm start} = 0.02$, $p_e^{\rm limit} = 1$, 
$p_f^{\rm start} = 1000$, $p_f^{\rm limit} = 1$, 
$p_v^{\rm start} = 0.02$, $p_v^{\rm limit} = 1$.
At the beginning of training, the force prefactor dominates the loss function, and the DP is mainly trained to minimise the force error. 
At the end of training the force prefactor decreases, while the energy and virial prefactors increase, so the training is in more balanced on minimising the energy, force and virial errors. 
It was argued that this dynamic prefactor scheme is usually more efficient than constant prefactors~\cite{Zhang_2018_prl}.
The model trained with this setting is usually very accurate in predicting atomic forces, but  is not always satisfactory in predicting the energy and virial tensor. 

Model fine-tuning restarts training.
In the fine-tuning, the model parameter are initialised from the DP trained with the default setting, the start learning rate is decreased to $10^{-4}$ while the stop learning rate is kept, and the loss prefactors set to
$p_e^{\rm start} = 10$, $p_e^{\rm limit} = 100$, 
$p_f^{\rm start} = 1$, $p_f^{\rm limit} = 1$, 
$p_v^{\rm start} = 10$, $p_v^{\rm limit} = 100$.
The idea behind this setting is to inherit model parameters from a DP with high force accuracy, and to train the model with significantly larger energy and virial prefactors to focus the training on the energy and virial accuracy. 
The  learning rate is decreased to slow the forgetting of the model parameters of the original model and to preserve the force accuracy. 
If the numerical error in the labels is low, then the fine-tuning usually leads to a significant improvement on the energy and virial accuracy with little decrement to the force accuracy~\cite{Zhang_2021_prl}.

Another hyper-parameter critical to the accuracy of the DP is the number of training steps.
Different settings may be employed for different DP purposes.
In the concurrent learning procedure, (for example DP-GEN see Sect.~\ref{sec:sec2c3}), 0.4-2 million   training steps should suffice.
If the DP  is intended for high-accuracy atomistic simulations, over 8 million training steps  are often employed for from-scratch training or fine-tuning.

Finally, the \lstinline{atom_ener} parameter may be used to specify the energy of an atom  in the vacuum for each atom. 
For metals and alloys, the cohesive energy calculated from DFT  does not always match  experiment. 
For example, for Ti, the cohesive energy calculated from DFT is 5.34 eV/atom~\cite{Schimka_2013_prb}, while experiment gives 4.85 eV/atom~\cite{Kittel_2005_book}. 
Because DP is fitted to DFT data, this command helps to correct the DFT ``errors" for the isolated atom. 
Accurate cohesive energy relative to experiment is critical for DP in some applications, such as the fracture behaviour of metals and alloys.

\paragraph{Interfacing with third-party packages.} 
After a DP is trained and tested by DeePMD-kit, application requires linking with other software.
DeePMD-kit provides Python and C++ interfaces for model inference, which is helpful for calculating  energy, atomic forces, and virial tensors with input atomic coordinates and the cell tensor.
DeePMD-kit also interfaces with ASE~\cite{Larsen_2017_jpcm}, LAMMPS~\cite{Plimpton_1995_jcp}, i-PI~\cite{Ceriotti_2014_cpc}, and GROMACS~\cite{Spoel_2005_jcc} for DP-based atomistic simulations. 
The detailed use of DP for these  can be found at~\footnote{\url{https://docs.deepmodeling.org/projects/deepmd/en/master/third-party/index.html}}.

Here, we explicitly discuss using  DPs in LAMMPS. 
The installation package for DeePMD-kit already incorporates the latest  stable version of LAMMPS.
Running MD  or other molecular simulations with DP in LAMMPS is very simple.
The user only needs to add two lines of commands specifying the interatomic interaction with the LAMMPS command \lstinline{pair_style} and \lstinline{pair_coeff} to the LAMMPS input script:
\begin{lstlisting}[
language=bash]
pair_style     deepmd graph.pb
pair_coeff     * *
\end{lstlisting}
where \lstinline{graph.pb} is the DP file. 
The user need not change any other part of the input script to conduct DP-based molecular simulations in LAMMPS.

\subsubsection{\label{sec:sec2c3}Deep Potential Generator} 

Deep Potential GENerator (DP-GEN) is a software package that implements the concurrent learning framework for generating high quality DPs.

\paragraph{Basic concepts and framework.} 
As  discussed in Sect.~\ref{sec:sec2b1}, generating a complete and compact dataset is critical for training high quality DPs. 
A straightforward way to generate data is to run finite-temperature \emph{ab initio} MD (AIMD) simulations  and use the configurations along the trajectories with labels (energy, forces, and virial tensors) as training datasets.
However, this method of data generation is not particularly efficient.
First, AIMD simulations are computationally expensive since  labels are calculated at each time step by DFT. 
Second, computationally affordable AIMD simulations are typically  very short and do not effectively explore phase space. 
Many important phenomena, such as phase transformations are difficult to be observed on AIMD time scales. 
Also configurations generated at successive MD time steps are extremely similar to one the other.

The DP-GEN concurrent learning scheme is designed to overcome  these difficulties by generating a much more complete and compact training dataset for DP training.
The DP-GEN is an iterative scheme. 
In each iteration it performs exploration, labelling and training. 
The training dataset is gradually enriched, and the quality of the DP  improves on each iteration. 
The DP-GEN scheme is deemed converged when all relevant configurations are explored and the DP is uniformly accurate on the explored configurations. 
The three steps in each DP-GEN iteration are as follows. 

\begin{figure}[]
	\includegraphics[width=0.48\textwidth]{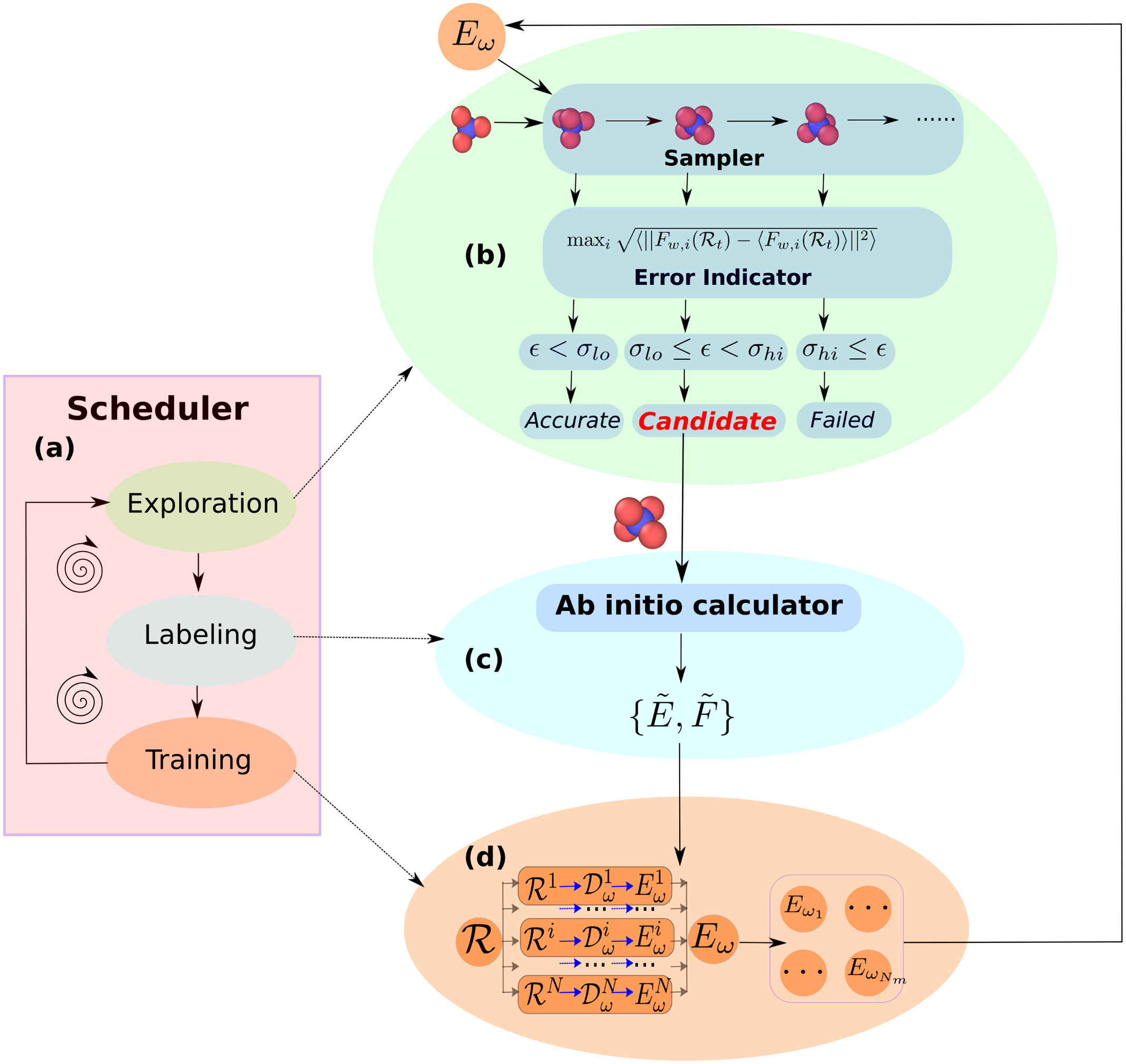}
	\caption{\label{fig:fig4}Schematic illustration of the DP-GEN scheme. (a) The DP-GEN scheduler runs iteratively, performing three steps: exploration, labelling, and training. (b) In the exploration step, different structures are sampled using DP MD and an error indication is applied to choose the candidates for labelling. (c) In the labelling step, DFT calculations are performed on the candidates to obtain the energy, forces, and virial tensor. (d) In the training process, DeePMD-kit package is used to train a new DP based on the initial datasets and candidates from each iteration. (Figure from~\cite{Zhang_2020_cpc}, with permission.)
	}
\end{figure}

(i) Exploration. 
We start with an ensemble of DPs. 
One of these DPs is used in the exploration step (Fig.~\ref{fig:fig4}(b)) to efficiently sample the relevant configuration space. 
The sampler typically performs several DP MD simulations (different initial configurations) for a set of thermodynamics conditions.  
In principle, sampling can also be performed using Monte Carlo simulations, enhanced sampling MD simulations~\cite{Yang_2021_prl}, and any molecular simulation method that explores  configuration space.

The machine learning potential models cannot be ``extrapolated" to the configurations that they are not trained on~\cite{Bartok_2018_prx}, thus the sampling should generate training datasets that explore the relevant configuration space as completely as possible.
In the context of the DP-GEN scheme, the configurations in the training dataset are proposed in the exploration step, thus it should explore the relevant configurations as complete as possible by exploiting the high efficiency of the DP itself. 
The design of the exploration strategy  depends on the applications for which the DP is intended. 
For example, a DP for liquid water properties need not fully explore ice configurations. 
A DP for chemical reactions may need enhanced sampling techniques to explore  reaction pathways usually not available by standard MD simulations~\cite{Yang_2021_ct}.

For each explored configuration, the difference between the DP prediction and the ground truth (i.e., error) is \emph{estimated} from the ensemble of models without referring to any DFT calculation. 
The error indicator is defined as the maximal deviation of the forces predicted by the ensemble of models. More precisely, the model deviation is
\begin{eqnarray}
	\epsilon_t=\mathop{max}_i\sqrt{\langle||\bm{F}_{w,i}(\mathcal{R}_t)-\langle \bm{F}_{w,i}(\mathcal{R}_t) \rangle||^2\rangle}, \label{equ31}
\end{eqnarray}         
where $\bm{F}_{\omega,i}(\mathcal{R}_t)$ is the force on atom $i$ predicted by model $E_w$ and the ensemble average $\langle \cdots \rangle$ is taken over the ensemble of models. 
The ensemble average $\langle \bm{F}_{\omega,i}(\mathcal{R}_t) \rangle$ can be approximated as:
\begin{eqnarray}
	\langle \bm{F}_{\omega,i}(\mathcal{R}_t) \rangle = \frac{1}{N_m}\sum_{\alpha=1}^{N_m}\bm{F}_{\omega_{\alpha},i}(\mathcal{R}_t), \label{equ32}
\end{eqnarray}  
where $N_m$ is the number of models in the ensemble (usually 4). 
Calculation of the model deviation  requires $N_m$ model inferences. 
Such an error estimate (over all explored configurations) provides an efficient measure of the model deviation for each configuration. 
Calculating ``real'' errors by comparing with DFT calculations would be extremely prohibitive.

\begin{figure}[]
	\includegraphics[width=0.48\textwidth]{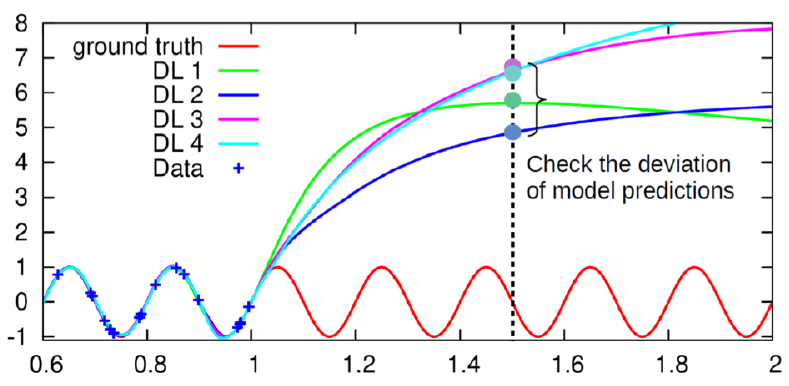}
	\caption{\label{fig:fig5}Visualisation of the model deviation. DL 1 to 4 indicates 4 Deep Learning models trained from the existing data (blue crosses).}
\end{figure}

How the model deviation indicates the error is shown in the schematic Fig.~\ref{fig:fig5}.
Four Deep Learning (DL) models (potentials), DL 1 to 4, are trained from the same training data (blue crosses) with random initialisation of the model parameters. 
At times, DL1 to DL4 all show small training error in the region ``covered" by the training data, but the predicting error is large in the region far away from the training data.
The standard deviation of the DL model predictions follows exactly the same trend.
In the region ``covered" by the training data, all  models/potentials are trained against the same target, thus their predictions agree with each other, while in the region far from the dataset the  different DL potentials (training from different initialisations  lead to different potentials) yield different predictions. 
Therefore, the model deviation is a good indicator of the DL prediction error.

Upper and lower trust levels, $\sigma_{hi}$ and $\sigma_{lo}$, are used to select configurations for labelling. 
When the error indicator ($\epsilon$) is smaller than $\sigma_{lo}$, it suggests that the atomic forces of all atoms in the configuration are accurately predicted and there is no need to do DFT labelling for this configuration. 
When $\epsilon \geq \sigma_{hi}$, the potential  behaves  poorly and DFT labelling will not be performed, because such explored configurations may be nonphysical (e.g.~overlapping atoms) due to poor potential quality.
Only when $\sigma_{lo} \leq \epsilon < \sigma_{hi}$, will the configuration  be selected as candidates for DFT labelling. 
The user may also set the maximal number of structures in the candidate set to perform DFT labelling; this leads to a randomly sampling to form the labeled dataset.

(ii) Labelling. In the labelling step (Fig.~\ref{fig:fig4}(c)), a DFT software package is called to calculate the energy, force and virial tensor on the selected configurations. 
The labeled data are \emph{added} to the existing training dataset.

(iii) Training. In the training step (Fig.~\ref{fig:fig4}(d)), an ensemble of new potentials is trained based on the same training dataset but with different and random initialisation of the potential parameters. 
Relatively short trainings are performed at this step, because the configurations are selected for labelling by the force criteria, and reasonably accurate DP  forces  can be obtained with relatively short trainings. 
Experience shows that 400,000 to 2,000,000  training steps are appropriate. 
One may also initialise the potential parameters in the  ensemble of potentials from the previous iteration, and start the training in the next iteration with a relatively small initial learning rate (e.g.,  $10^{-4}$), and with a bias to the new training data.

(iv) Initial datasets for starting the DP-GEN workflow. The DP-GEN workflow introduced above is automatic, but human intervention is required at the beginning when no DP exists. 
Generally, we generate some initial datasets and train an ensemble of initial DPs to kick off the workflow. 
In metals and alloys, the generation of initial bulk datasets usually consists of the following steps. 
(a) Start from a supercell ($2 \times 2 \times 2$) of crystal unit cells (BCC, FCC, HCP,...). 
(b) Compress and dilate the supercell uniformly to cover a range of densities. 
(c) Randomly perturb the atom coordinates and supercell vectors. 
(d) Randomly replace atom elements to generate alloys. 
(e) Start from different structures and do a short AIMD at low temperature (100 K). 
(f) Collect the DFT results and generate initial training datasets. 
The initial datasets are not particularly important as long as they reasonably cover the starting structures of DP MD in the exploration step to avoid breakdown of the DP MD within the first few steps. 
This generating initial dataset method  is simply one possibility;  other methods such as  generating initial structures according to crystal symmetry in genetic algorithm~\cite{Glass_2006_cpc} are also applicable.


\paragraph{A practical guide}
Here we provide a practical guide for generating a general-purpose DP for a bulk metal or alloy using the DP-GEN scheme.

(i) Generation of initial datasets. Initial datasets are needed to generate the first ensemble of DPs. 
Because initial datasets are only used to start DP-GEN, datasets with only several hundred different configurations should suffice.
The accuracy of the DPs is not sensitive to the choice of the initial datasets and would be gradually improved as iterations increase  assuming that  the   DP-GEN run parameters are set appropriately.

(ii) Exploration. The exploration strategy is not unique and is determined by the user according to the applications of interest. 
Here, we  focus on the general exploration strategy using, again,   bulk metals or alloys as our example. 
DP-based MD simulations at different temperatures and pressures are employed to explore the configurational space. 
For efficient sampling,   we increase the temperature during the exploration. 
If the highest melting point of the element in an alloy is $T_m$, we usually divide the temperature range from 50 K to 2$T_m$ into 4 regions. (a) [50 K, (0.1 , 0.2 , 0.3 , 0.4)$T_m$] (b) [0.5, 0.6, 0.7, 0.8, 0.9]$T_m$; (c) [1.0, 1.1, 1.2, 1.3, 1.4]$T_m$; (d) [1.5, 1.6, 1.7, 1.8, 1.9]$T_m$. 
For each temperature region, the pressure range is varied over [0.001, 0.01, 0.1, 1, 5, 10, 20, 50] kBar. 
In this case, there are $5 \times 8=40$ different MD conditions in each temperature region. 
Other parameters for the DP MD simulations is the MD simulation duration and the number of different initial structures. 
At the beginning of each new temperature region, the number of MD step should be small (500 or 1000) and the number of starting structures should also be small (e.g., 5 each for different distorted crystal supercells). 
This is because  the DPs have little ``knowledge'' of this new temperature region such that  the expected accuracy is low. 
As the iteration number increases within a temperature region, the MD runs are longer and the number of starting structures is increased. 

After the MD runs, candidate systems are selected based upon the trust levels $\sigma_{lo}$ and $\sigma_{hi}$. 
The key idea is to make DP-GEN converge within each temperature region. 
If $\sigma_{lo}$ and $\sigma_{hi}$ are too strict, for example $\sigma_{lo}$ is too low, the accurate ratio ($\epsilon < \sigma_{lo}$) stays low (smaller than 90\%) in the temperature region. 
If $\sigma_{hi}$ is too high, too many structures are selected as candidates for labelling and the ratio of the failed structure ($\epsilon \geq \sigma_{hi}$) is very low. 
The values of $\sigma_{lo}$ and $\sigma_{hi}$ used in our previous work for different systems (see  Table~\ref{tab:table2});  $\sigma_{hi}$ is normally 0.15-0.30 eV/{\AA} higher than $\sigma_{lo}$.
In principle, $\sigma_{lo}$ should be higher than the lowest error that a DP can achieve (error introduced by data, fitting ability of optimiser, representability of the DP, \ldots). 
$\sigma_{hi}$ should be set so as to eliminate  unphysical configurations from the dataset, since the convergence of the DFT calculations for such configurations is difficult to obtain.
See~\cite{Zhang_2020_cpc} for additional guidance.

\begin{table}[]
	\caption{\label{tab:table2} Trust levels $\sigma_{lo}$ and $\sigma_{hi}$ employed in DP-GEN for several systems.}
	\begin{ruledtabular}
		\begin{tabular}{cc}
			\textrm{System}&[$\sigma_{lo}$,$\sigma_{hi}$]\\
			\colrule
			Mg~\cite{Zhang_2019_prm} & [0.03,0.13]\\
			Al~\cite{Zhang_2019_prm} \& Al-Mg~\cite{Zhang_2019_prm} & [0.05,0.15]\\
			Cu~\cite{Zhang_2020_cpc} & [0.05,0.20]\\
			Mg-Al-Cu~\cite{Jiang_2021_cpb} & [0.05,0.20]\\
			Ti~\cite{Wen_2021_npj} & [0.10,0.25] at $T < 1.5T_m$\footnote{$T_m$ is 1941 K, which is the experimental melting point for Ti.} for bulk\\
			& exploration and [0.15,0.30] elsewhere\\ 
			W~\cite{Wang_2021_Tungsten_arxiv} & [0.20,0.35]\\
			Ag-Au~\cite{Wang_2021_msmse} & [0.05,0.20]\\
			water~\cite{Zhang_2021_prl} & [0.15,0.25] in first 24 iterations\\
			& [0.18,0.32] in iterations 25 to 32\\
			& [0.20,0.35] in iterations 33 to 36\\
			SiC~\cite{Fu_2021_arxiv} & [0.15,0.30]\\
			Li$_{10}$(Ge,Si, or Sn)P$_{2}$S$_{12}$~\cite{Huang_2021_jcp} & [0.12,0.25]\\
		\end{tabular}
	\end{ruledtabular}
\end{table}      

(iii) Labelling. After  candidate structures are selected from the exploration step, users can choose the maximal number of structures sent to DFT labelling in each iteration.
On the one hand, if there is a large number of structures for DFT labelling in one iteration, the computational cost for labelling would be high and this dataset of configurations may be redundant, leading to wasted computational resources. 
On the other hand, if the number of structures for DFT labelling is too small (e.g., one) in one iteration, the datasets after DP-GEN run would be very compact, requiring many iterations, which again would be computationally expensive for training in each iteration. 
In this case, the maximal number of structures for DFT labelling in one iteration should be set appropriately; we typically set this number to $\sim$100.

(iv) Training. The prefactor of the virial tensor in the loss function (Eq.~\eqref{eq:loss}) is usually set to zero in the DP-GEN training loop for  two reasons.
First, only the model deviation of the forces is used to select candidates in the exploration step; this is irrelevant for the virial tensors.
Second,  longer training is needed to obtain a DP with good accuracy on both forces and virial tensors. The training step in each iteration of DP-GEN should be small, so we do not initially train on the virial tensors.

(v) DP-GEN convergence. In principle, DP-GEN is considered converged when the model deviation ($\epsilon$) on each structure is smaller than the trust level $\sigma_{lo}$.
However, there are often  some (very low probability) configurations with undesirable accuracy  in long enough MD simulations. 
These  configurations generally do not improve the DP performance. 
In this case, DP-GEN allows users to set the convergence criterion at which level the DP is regarded as converged.
For example, the user can skip the labelling step when the percentage of configurations with model deviation smaller than $\sigma_{lo}$ is larger than 99\%.
After convergence,  long training steps (usually over 8,000,000 steps) are performed on all DP-GEN loop datasets   by including labelled energy, forces, and virial tensors. 
Comprehensive tests of the resulting model are then performed and the Autotest in DP-GEN package is applied to perform  different property tests (see below). 
The DP obtained after careful DP-GEN run and long training is normally  good for general purposes, but does not guarantee good performance for subtle properties. 
Specialisation for these properties is required and the details of specialisation are presented in Sect.~\ref{sec:sec2e2}. 

\paragraph{Autotest.} After one or a series of DPs is trained, we employ the Autotest package, to calculate a simple set of properties and/or perform tests for comparison with DFT and/or empirical interatomic potentials (EAM, MEAM, and etc.). 
Because the DP was obtained by fitting DFT calculation results, the degree of agreement between DP and DFT for a series of properties is usually excellent; if not, this comparison provides a guide for further training  and parameter settings. 
Autotest, as part of the DP-GEN package, has standardised the calculation of some critical physical properties and provides a series of reliable benchmark testing to better evaluate the performance of the DPs. 
As of this writing, Autotest includes the calculation of the following set of properties (additional properties are added continuously):
(a) equilibrium structural parameters (relaxation),
(b) equation of state (eos),
(c) elastic constants (elastic),
(d) vacancy formation energy (vacancy),
(e) interstitial formation energy (interstitial), and
(f) surface formation energy (surface).

Autotest can also use LAMMPS or VASP to reproduce and refine previous calculation results. The current package is mainly targeted for simple crystal structures (metals and alloys), but is extensible for users to implement new features. The detailed manual and framework of Autotest package is available at \footnote{\url{https://github.com/deepmodeling/dpgen/wiki}}.

\subsection{\label{sec:sec2d}The DP Library}
\begin{figure}[!htbp]
	\includegraphics[width=0.48\textwidth]{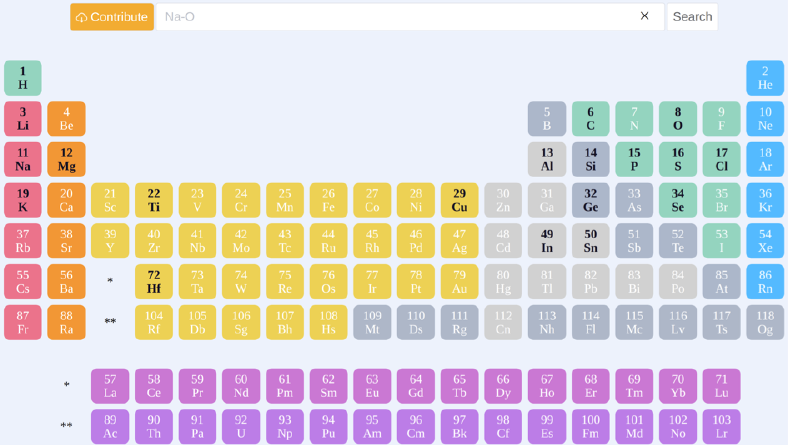}
	\caption{\label{fig:fig6} Elements for which DPs are currently available in the DP Library \url{https://dplibrary.deepmd.net} are indicated in black.}
\end{figure}

The DP-GEN package provides a relatively automatic routine for generating DPs, and the DP Library project is a place for sharing and publishing the DPs and the training datasets.
First, DP Library is a place for model sharing, which is similar to other interatomic potential repositories like the NIST~\footnote{https://www.ctcms.nist.gov/potentials/} and OpenKIM~\footnote{https://openkim.org/}. 
Second, DP Library provides an opportunity for data sharing.
One can contribute and download DFT datasets used to train a published DP. 
If a DP needs to be refined or improved, one can first add new data to the downloaded training datasets from DP Library and then perform training using DeePMD-kit. 
For example, one may develop a DP for A-B alloys based on the training data for element A and element B, instead of generating the datasets by him/herself.
The settings (usually settings to use a DFT software) used to generate the dataset are asked to provide, so the new data can be generated in the same way as the downloaded dataset.
The shared training data makes the published DPs reproducible and improvable. 
DP Library can be accessed at~\footnote{\url{https://dplibrary.deepmd.net}} and an image of the website is shown in Fig.~\ref{fig:fig6}. 
In the periodic table, the available DP for elements is in black and details about the DP and DFT data can be found by just several clicks. For more details, please refer to the manual of DP Library at~\footnote{\url{https://dplibrary.deepmd.net/#/help}}.

\subsection{\label{sec:sec2e}Efficiency and Accuracy of DPs for Applications}

To make DP more practical for applications in different materials systems, efficiency and accuracy must be balanced. 

Although DP is very much faster than DFT, 
it is still much slower than empirical interatomic potentials like EAM and MEAM. 
This is not surprising given the  vast number of parameters in the neural net of the DP (compared to simple empirical potentials such as EAM).
Hence, optimisation of the efficiency in use of the DP can greatly enhance the practical usability.
Recently, a highly optimised GPU version of DeePMD-kit pushed the limit of MD with \textit{ab initio} accuracy to 100 million atoms~\cite{Lu_2021_cpc,Jia_2020_sc20} (recognised by the 2020 Gordon Bell Prize). 
Further optimisations are still possible based on advances in neural networks.

DPs generated by the example exploration protocol (Sect.~\ref{sec:sec2c3}) may not be accurate for all applications. 
For example, the Ti DP yields screw $\langle \bold{a} \rangle$ dislocation properties that are not consistent with experiment (or DFT).
For some particular applications, the DP must be tuned or specialised for important, yet subtle properties. This may be accomplished through specialisation, as discussed below~\cite{Wen_2021_npj}.
 
We first introduce DP model compression which can easily accelerate DP by a factor of 4 to 18, based on experience. 
Next, we discuss DP specialisation and consider the example of Ti, mentioned above.

\subsubsection{\label{sec:sec2e1}DP compression}

The most computationally intensive part of using the DP is the evaluation of the embedding net (Eq.~\eqref{equ23} or \eqref{equ24}) and the assembly of the descriptor by  Eq.~\eqref{equ26}--\eqref{equ28}~\cite{Lu_2021_arxiv}. 
The goal of  DP compression is to reduce the computational and memory overhead associated with the embedding net with little loss of accuracy. 
The evaluation of the DP involved the mapping from a scalar to a vector of dimension $M$, each dimension of which can be approximated by a piece-wise fifth order interpolating polynomials. 
The range of the embedding net is first discretised by nodes $x_1, x_2, \dots, x_l, \dots, x_{L+1}$, and the lengths of intervals (tabulation step) are assumed to be the same. 
On each interval, e.g., $[x_l, x_{l+1})$, the 6 fitting parameters of the fifth-order polynomials are uniquely determined by matching the value and the first and second derivatives of the polynomial to those of the embedding net on nodes $x_l$ and $x_{l+1}$.
This approximation is referred to as \emph{tabulation}, because the parameters of the polynomials are stored in a table, and use involves simply a look-up the table for the polynomial when evaluating the embedding net. 
The accuracy of the tabulation is controlled by the tabulation step; a size of 0.01 leads to errors in energy and force smaller than $10^{-7}$~eV/atom and $10^{-6}$~eV/\AA. 
After tabulation, the multiplication between the embedding and the environment matrices becomes the bottleneck.
The embedding matrix is stored  in the memory after it is computed and then  loaded from  memory to the register for the matrix multiplication.
This requires significant I/O with the   memory.
This can be eliminated by merging the tabulation and the multiplication; i.e., once one component of the embedding net is computed by tabulation, it is immediately multiplied with the environment matrix pre-loaded in the register, and accumulated to the result. 
With this optimisation, the multiplication between the embedding net and the redundant zeros in the environment matrix can be avoided. 

The DP compression was  benchmarked for Cu, H$_2$O and Al-Cu-Mg ternary alloy DPs~\cite{Lu_2021_arxiv}. 
The model inference was accelerated by 9.7, 4.3 and 18.0 times on a CPU and by 9.7, 3.7 and 16.2 times on an Nvidia V100 GPU for Cu, H$_2$O and Al-Cu-Mg, respectively.
The maximum number of atoms handled by one GPU also increased from 12, 49 and 5 $\times 10^3$  to 129, 246 and 61 $\times 10^3$, respectively.

\begin{figure}[]
	\includegraphics[width=0.45\textwidth]{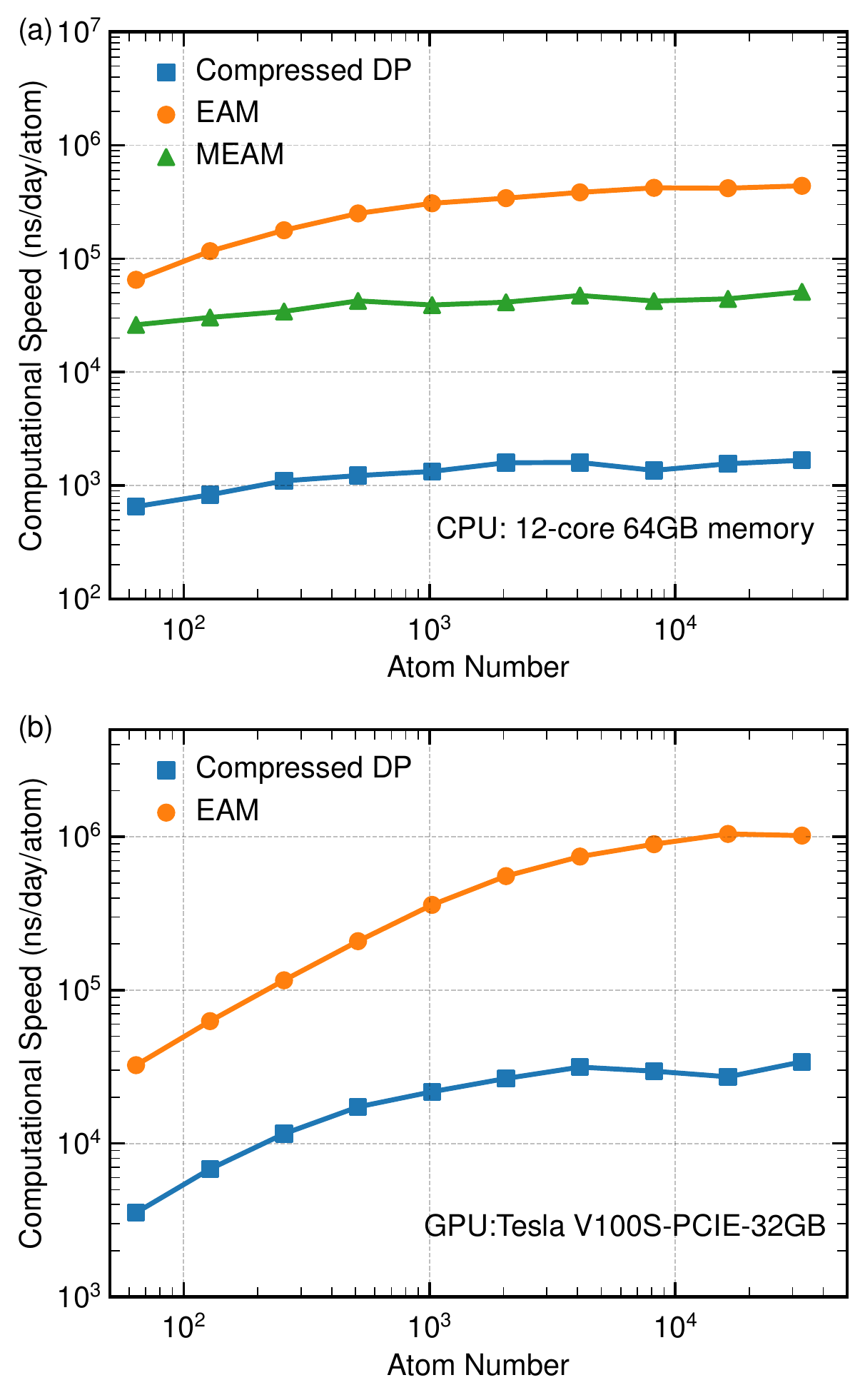}
	\caption{\label{fig:fig7} A comparison of the speed of MD simulations using  the compressed Ti DP, an EAM, and/or an MEAM potential on (a) CPU and (b) GPU systems~\cite{Wen_2021_npj}.
	}
\end{figure}

Figures~\ref{fig:fig7}(a) and (b) display the speed comparison of the compressed Ti DP~\cite{Wen_2021_npj} with an EAM~\cite{Mendelev_2016_jcp}, and an MEAM potential~\cite{Hennig_2008_prb} on a CPU and GPU machine. 
Note that the DP has a larger radius cutoff distance than EAM and MEAM in this case. 
On CPUs, the compressed DP is 200-300 times slower than EAM potentials and 30-40 times slower than the MEAM potential. 
On GPUs, the compressed DP is 20-30 times slower than the EAM potential (MEAM is currently not ported to GPU in LAMMPS). 
All potentials show a linear scaling with the number of atoms. 
Because of this linearity and speed, the compressed DP can be used to perform large scale MD simulations to investigate a wide range of properties with \textit{ab inito} accuracy; e.g., defect properties, phase transformations, and transport coefficients. 

DP compression is supported in the DeePMD-kit package (releases beyond 2.0.0) and the compressed model can be easily generated using the following command:
\begin{lstlisting}[
language=bash]
dp compress -i graph.pb -o graph-compress.pb  
\end{lstlisting}
Using the optional \lstinline{-s} flag, followed by the tabulation step with the default value of 10$^{-2}$ typically gives very accurate compressed DPs (compared with the original). 
It is recommended that the user compare the values of a few key properties from the original and compressed DPs.

\subsubsection{\label{sec:sec2e2}Specialisation}

\begin{figure*}[!htbp]
	\includegraphics[width=1.0\textwidth]{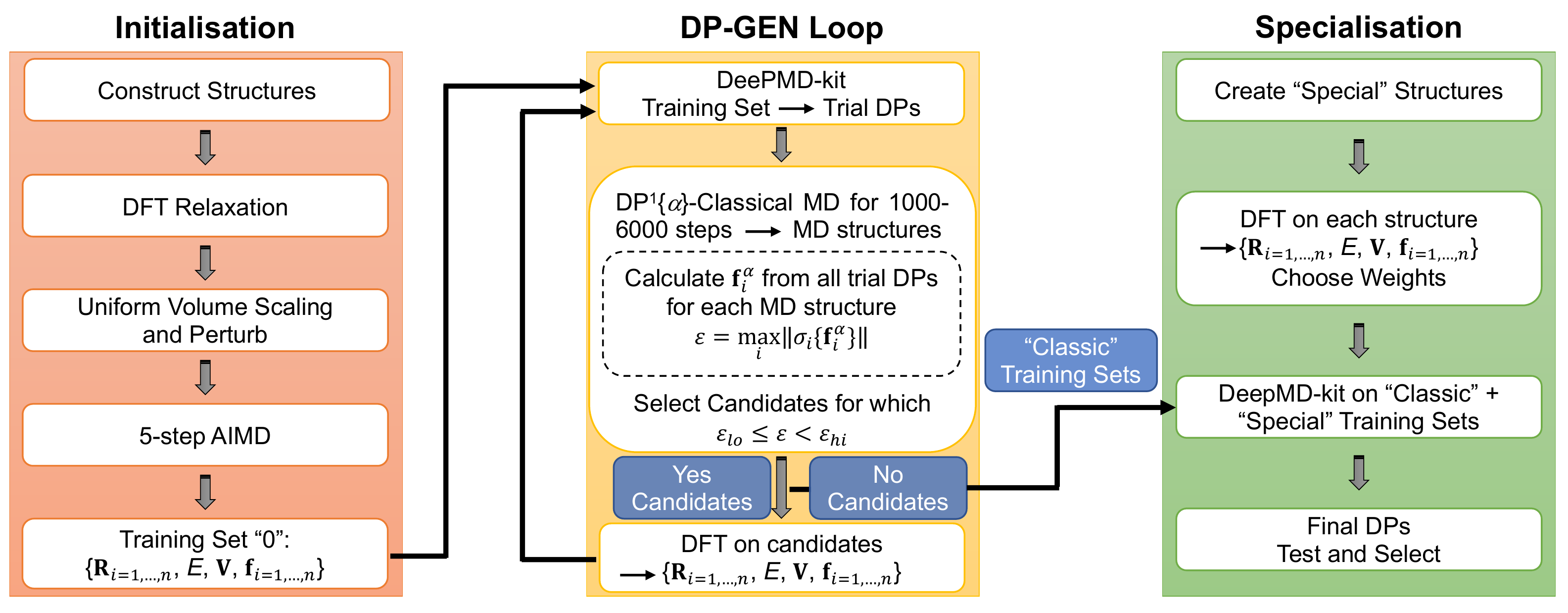}
	\caption{\label{fig:fig8}The workflow for specialising a general DP. $\mathbf{R}_i$ is the atomic coordinate of atom $i$, $E$ is the total energy of one configuration, $\mathbf{V}$ is the virial (stress) tensor of one configuration, $\mathbf{f}_i$ is the force on atom $i$, $n$ is the number of atoms in one configuration, DP$^1$ is the first ensemble of trial DPs, $\alpha$ labels the $\alpha^\text{th}$ DP in the ensemble, $\sigma$ is the standard deviation, and $\epsilon_\text{lo}$ and $\epsilon_\text{hi}$ are two thresholds in DP-GEN. The specialisation step is shown in the green box. Figure from~\cite{Wen_2021_npj}, \href{https://creativecommons.org/licenses/by/4.0/}{CC BY 4.0}. 
	See~\cite{Wen_2021_npj} for details.}
\end{figure*}

The DP generated from the DP-GEN scheme  described above may not be sufficiently accurate  for some complex phenomena and specialisation of this general-purpose DP may be required. 
This is not surprising since the exploration strategy may not provide a sufficient sampling of the relevant local structures that are inherent to the complex phenomena of interest. 
Therefore, some special structures should be added to the training process to better represent the requisite subtle properties.
The workflow for specialising DP is shown in Fig.~\ref{fig:fig8}. The initialisation and DP-GEN loop steps were  discussed in Section~\ref{sec:sec2c3}. 

In the specialisation step, ``Special" structures are first created based on the special properties/structures of interest. 
For example, if the DP elastic constants are not sufficiently accurate, ``Special" structures may be created corresponding to a range of different crystal deformations. 
If the specialised DP is to describe dislocation properties in complex crystal, then ``Special" structures that include sheared structures akin to those used in determining the generalised stacking fault energy $\gamma$-lines~\cite{Vitek_1968_pma} may be of use. 
DFT calculations are then performed on these ``Special" structures and the energies, atomic forces, and virial tensors along with the configuration form the ``Special" training sets. 
The ``Special" training sets then combine with the ``Classic" training sets from DP-GEN loop and weights are chosen for the ``Special" training sets. 
The default weight is 1 but this weighting of the  ``Special" training sets should  be increased because 
(i) the number of ``Special" training sets are usually much smaller than those in the ``Classic" training sets, and 
(ii) experience shows this may be necessary to properly reproduce the special properties. 
Next, DeePMD-kit package is applied to retrain the DP on all of the training sets.
Finally, the specialised DP is further tested to insure the best overall performance.

\begin{figure*}[!htbp]
	\includegraphics[width=1.0\textwidth]{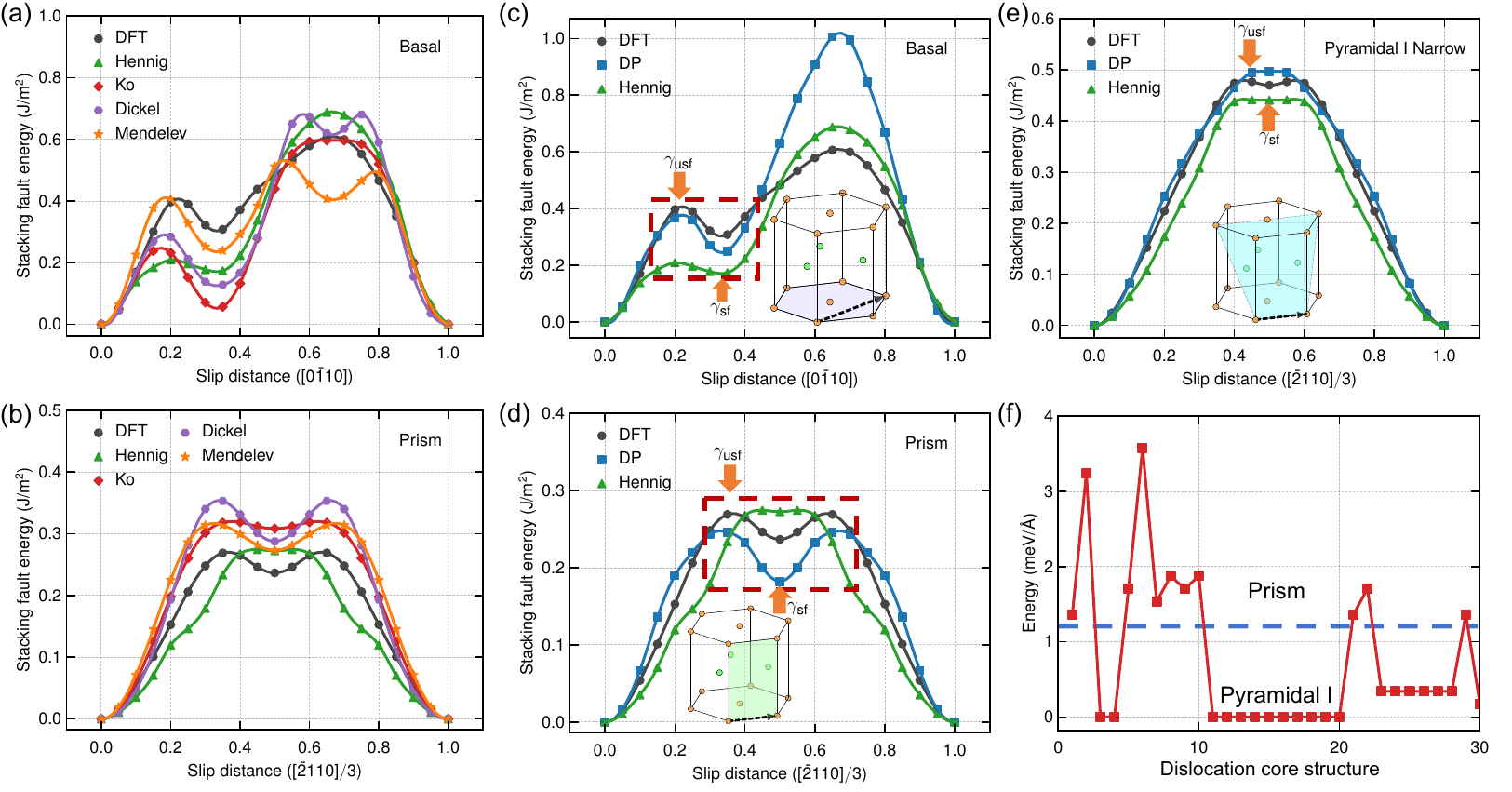}
	\caption{\label{fig:fig9}The generalised stacking fault energy ($\gamma$-lines) on the (a,c) Basal, (b,d) Prism, and (e) Pyramidal I plane of HCP Ti calculated using DFT, empirical potentials (Hennig~\cite{Hennig_2008_prb}, Ko~\cite{Ko_2015_prb}, Dickel~\cite{Dickel_2018_msmse}, and Mendelev~\cite{Mendelev_2016_jcp}), and a specialised DP. The configurations in the dashed red box and at zero slip (origin) are included in the training sets. All configurations on the Pyramidal I narrow $\gamma$-line (e) are included in the training dataset. 
(f) Each point denotes the energy of a simulation cell containing a screw $\langle \mathbf{a} \rangle$ dislocation core from a 600 K MD simulation after quenches to zero temperature (each point is one picosecond apart in the MD simulation). Two types of screw dislocation cores are observed: one delocalised onto a Prism plane (above the blue line) and one on Pyramidal I plane (below the blue line).  Some of the cores are slightly distorted  leading to small variations in the energy. 
See~\cite{Wen_2021_npj} for details.
	}
\end{figure*}

Here, we show an example of DP specialisation for the mechanical response of Ti~\cite{Wen_2021_npj}. 
Figs.~\ref{fig:fig9}(a) and (b) present a comparison of different empirical interatomic potentials with DFT on the $\gamma$-lines of Basal and Prism planes in HCP Ti. 
Illustrations of the Basal and Prism planes of HCP Ti are shown in the insets of Figs.~\ref{fig:fig9}(c) and (d). 
All of these empirical potentials yield inaccurate $\gamma$-line profiles and/or generally predict low stacking fault energy (the first minimum at $\sim$0.33 along the $x$-axis) on the Basal plane with respect to DFT. 
These empirical potentials are not specialised for dislocation properties.
When general-purpose DP is obtained from DP-GEN scheme, it also shows  similar systematic inadequacy as the other empirical potentials: low Basal and high Prism stacking fault energy, which prohibits the use of a generalised DP for investigating dislocation properties in HCP Ti. 
Specialisation is then performed by adding the structures in the red dashed box as well as the origin in Figs.~\ref{fig:fig9}(c), (d) and the structures along $\gamma$-lines in Fig.~\ref{fig:fig9}(e), all with a weight of 100.
The DP $\gamma$-lines in Figs.~\ref{fig:fig9}(c)-(e) are overall in good agreement with DFT. 
In particular, the stable stacking fault energy order Basal $>$ Prism $>$ Pyramidal I narrow (the stable stacking fault energy on this plane will decrease significantly after full relaxation compared to that from $\gamma$-line) is different from previous empirical potentials in Figs.~\ref{fig:fig9}(a) and (b) and follows the trends observed in the DFT data. 
Based on this, the relative energies of the screw $\langle \mathbf{a} \rangle$ dislocation on the Prism and Pyramidal I planes are shown in Fig.~\ref{fig:fig9}(f).
The screw $\langle \mathbf{a} \rangle$ dislocation is more stable on Pyramidal I plane than on Prism plane for DP, which agrees with previous DFT calculations and experimental measurements~\cite{Clouet_2015_nm}. 
We refer readers to~\cite{Wen_2021_npj} for more details. 
This example demonstrates the importance of specialisation of the DP and the superior flexibility of DP over other empirical potentials.

\section{\label{sec:sec3}DP applications in materials science}

\begin{table}[!htpb]
	\caption{\label{tab:table3} Example applications of DP in materials science.}
	
	\begin{ruledtabular}
		\begin{tabular}{cc}
			\textrm{System}&Reference\\
			\colrule
			\textbf{Elemental Bulk Systems} &\\
			Al & \cite{Zhang_2019_prm,Wang_2019_apl,Zeng_2021_prr,Liu_2020_jpcm,Liu_2021_mre,Cheng_2021_aipadv}\\
			Mg & \cite{Zhang_2019_prm}\\
			Cu & \cite{Zhang_2020_cpc}\\
			Ti, W & \cite{Wen_2021_npj,Wang_2021_Tungsten_arxiv}\\
			Ag, Au & \cite{Andolina_2021_jpcc,Wang_2021_msmse,Chen_2021_arxiv}\\
			Li & \cite{Jiao_2021_arxiv}\\
			Be & \cite{Zhang_2020_pp}\\	
			Ga & \cite{Niu_2020_nc}\\
			Sb & \cite{Shi_2021_mssp}\\
			C & \cite{Wang_2022_carbon}\\
			Si &\cite{Bonati_2018_prl,Li_2021_mtp}\\
			P & \cite{Yang_2021_prl}\\
			
			\textbf{Multi-element Bulk Systems} &\\
			Al-Mg, Al-Cu-Mg & \cite{Zhang_2019_prm,Wang_2020_fc,Andolina_2021_prm,Jiang_2021_cpb}\\
			Al-Cu, Al-Zn-Mg & \cite{Bourgeois_2020_nc,Cheng_2020_scripta}\\
			Al-Cu-Ni & \cite{Ryltsev_2021_arxiv}\\
			Ag-Au & \cite{Andolina_2021_jpcc,Wang_2021_msmse}\\
			Pd-Si, Nb$_5$Si$_3$, Zr$_{77}$Rh$_{23}$, Bi$_2$Te$_3$ & \cite{Wen_2019_prb,Wang_2021_jap,Guo_2019_jac,Guo_2019_njp}\\
			Al$_{90}$X$_{10}$ (X=Tb, Cr, or Ce) & \cite{Tang_2020_pccp,Tang_2021_acta,Han_2020_prl,Tang_2021_jcp}\\
			(Pd, Pt)$_ \mathrm{x}$(Ge, Sn, Pb)$_ \mathrm{y}$ & \cite{Daniels_2021_jpcc}\\
			P$_2$Sn$_5$ & \cite{Zhang_2021_jpcc}\\
			silica, silicate & \cite{Balyakin_2021_pre,Deng_2021_grl,Luo_2021_grl,Luo_2021_gca}\\
			SiC & \cite{Fu_2021_arxiv,Chen_2021_jap}\\
			B$_4$C & \cite{An_2021_prm}\\
			molten salt LiF, FLiBe, and chloride & \cite{Rodriguez_2021_acsami,Liang_2021_jmst,Liang_2020_ats,Pan_2020_cms,Pan_2021_cms,Liang_2021_acsami,Bu_2021_semsc,Zhao_2021_ionics,An_2021_jpcb}\\
			Li or Na-based battery materials &\cite{Xu_2020_jpcc,Huang_2021_jcp,Marcolongo_2019_arxiv,Gupta_2021_ees,Li_2021_icf,Lin_2021_angew}\\
			TiO$_2$ & \cite{Andrade_2020_prm}\\
			$\beta$-Ga$_2$O$_3$ & \cite{Li_2021_apl}\\			
			ferroelectrics HfO$_2$ & \cite{Wu_2021_prb}\\			
			Ag$_2$S & \cite{Balyakin_2022_cms}\\
			MoS$_2$ & \cite{Wang_2020_arxiv}\\
			SnSe & \cite{Guo_2021_mte}\\
			Zr$_{1-x}$W$_x$B$_2$ & \cite{Dai_2020_jecs}\\
			(Hf$_{0.2}$Zr$_{0.2}$Ta$_{0.2}$Nb$_{0.2}$Ti$_{0.2}$)X (X=C or B$_2$) & \cite{Dai_2020_jmst,Dai_2021_jmst}\\
			
			&\\
			\textbf{Aqueous Systems} &\\
			water & \cite{Ko_2019_mp,Sommers_2020_pccp,Zhang_2020_prb,Gartner_2020_pnas,Andreani_2020_jpcl,Xu_2020_prb,Piaggi_2021_jctc,Zhang_2021_prl,Tisi_2021_arxiv,Zhang_2021_jpcb,Torres_2021_jpcb,Shi_2021_jpcl,Calio_2021_jacs}\\
			zinc ion in water & \cite{Xu_2019_jpca}\\
			water-vapor interface & \cite{Samuel_2021_jcp,Galib_2021_science}\\
			water-TiO$_2$ interface & \cite{Andrade_2020_cs}\\
			ice & \cite{Piaggi_2021_mp,Ye_2021_prl}\\
			&\\
			\textbf{Molecular Systems and Clusters} &\\
			organic molecules & \cite{Jiang_2021_arxiv,Zeng_2021_ef,Chen_2018_jpcl,Yang_2021_ct,Zhang_2018_jcp,Wang_2020_sm,Pan_2021_jctc}\\
			metal and alloy clusters & \cite{Tuo_2020_jcp,Andolina_2021_jpcc}\\
			&\\
			\textbf{Surfaces and Low-dimensional Systems} &\\
			metal and alloy surfaces & \cite{Andolina_2021_prm,Andolina_2021_jpcc,Wang_2021_msmse}\\
			graphane & \cite{Wang_2022_carbon,Achar_2021_jpcc}\\		
			monolayer In$_2$Se$_3$ & \cite{Wu_2021_prb2}\\
			2D Co-Fe-B & \cite{Chen_2021_an}\\

		\end{tabular}
	\end{ruledtabular}
\end{table} 

In the past three years, DPs have been applied in a number of systems in materials science including (i) elemental bulk systems, (ii) multi-element bulk systems, (iii) aqueous systems, (iv) molecular systems and clusters, and (v) surfaces and low-dimensional systems. 
Table~\ref{tab:table3} shows a  list of the material systems to which DPs  have been applied (as of the writing of this paper). 
We choose several examples from each category to briefly discuss the corresponding DP application and how DP  aids materials science research.

\subsection{\label{sec:sec3a}Elemental Bulk Systems}
To date, DPs have been applied to a wide-range of pure  systems, including Al, Mg, Cu, Ti, W, Ga, C, Si, \ldots~as shown in Table~\ref{tab:table3}. 
Al was the first metal system to  which DP was applied  and a  general-purpose  DP developed~\cite{Zhang_2019_prm}. 
It accurately reproduces the lattice parameter, elastic constants, vacancy and interstitial formation energies, surface  energies, twin and stacking fault energies, melting point, enthalpy of fusion,  diffusion coefficient, \ldots. 
The general-purpose DP for other metal elements is also accurate for the same properties. 
For other properties,  not included in the training datasets, DP is in better agreement with DFT than MEAM~\cite{Baskes_1992_prb} for phonon dispersion relations, equations of state, and the liquid state radial distribution function~\cite{Zhang_2019_prm}. 
Based on this, Wang et.~al.~\cite{Wang_2019_apl} smoothly interpolated the Ziegler-Biersack-Littmark (ZBL) screened nuclear repulsion potential with a DP to obtained a  DP-ZBL model for irradiation damage simulations, surpassing the widely adopted ZBL MEAM~\cite{Pascuet_2015_jnm} or EAM~\cite{Jacobsen_1987_prb} potentials. 
Later, a DP was developed for warm dense Al to simulate  ion dynamics near the hydrodynamic limit~\cite{Zeng_2021_prr}, structural and dynamic properties~\cite{Liu_2020_jpcm}, and electronic and ionic thermal conductivities~\cite{Liu_2021_mre}.
A DP was also developed for high temperature and high pressure liquid Al to calculate shear velocity~\cite{Cheng_2021_aipadv}. 

Dislocation properties play  important roles in the plastic response of most structural materials, including Ti and W. 
The specialised Ti DP accurately depicts the $\gamma$-lines on different planes (Figs.~\ref{fig:fig9}(c-e)) and the screw $\langle \mathbf{a} \rangle$ dislocation core energy ordering  between the Prism and Pyramidal I planes (Fig.~\ref{fig:fig9}(f)). 
In addition, the screw $\langle \mathbf{a} \rangle$ dislocation core structures on Prism and Pyramidal I planes are also in surprisingly good agreement with DFT results~\cite{Clouet_2015_nm}.
Because the dislocation core structures could not be explicitly included into DFT training datasets, the example of Ti DP shows that surrogate properties can be used to optimise DPs for dislocation properties of complex HCP systems.
While the dislocation core structures of BCC W are not as complex as in HCP Ti (because it only adopts a compact core), the Peierls barrier for BCC W is difficult to reproduce with other empirical potentials (cf. DFT results~\cite{Wang_2021_Tungsten_arxiv}). 
The DP-SE2 model with only a two-body embedding descriptor yields a very poor prediction on the Peierls barrier, but the DP-HYB that hybridises descriptors with two-body and three-body embeddings reproduces this property very accurately~\cite{Wang_2021_Tungsten_arxiv}.
The  Ti and W examples demonstrate that DPs can accurately describe dislocation core structures and Peierls barrier of BCC and HCP metals.

The DP approach has also been applied to Ag and Au (widely used in catalytic applications). 
Andolina, et al.~\cite{Andolina_2021_jpcc} and Wang, et al.~\cite{Wang_2021_msmse} developed DP for Ag and Au that is accurate at lattice parameters, elastic constants, surface formation energies, interstitial and vacancy formation energies, et al. 
Furthermore, Andolina, et al.~\cite{Andolina_2021_jpcc} got accurate adsorption energy and diffusion barriers for adatoms on \{100\}, \{110\}, and \{111\} compared to DFT results. 
Wang, et al.~\cite{Wang_2021_msmse} presented a comprehensive study of the Au \{111\} surface reconstruction  using a DP that yields excellent agreement with  DFT results. 
From another perspective, Chen, et al.~\cite{Chen_2021_arxiv} used DP MD to illustrate the dynamics compression process of Au. 
The developed DP could accurately reproduce the experimentally determined phase boundaries and the short-to-medium range orders are proposed to reduce the Gibbs free energies of the shocked structures.
The examples above for Ag and Au validate the applications of DP in both catalytic and shock compression areas.

In addition, DP has also been applied in many other elemental bulk systems and here we only list some of the examples in Table~\ref{tab:table3}. 
For Li, which is an important element for battery, Jiao, et al.~\cite{Jiao_2021_arxiv} developed DP to reveal self-healing mechanisms in a large Li-metal system. 
For Ga, Niu, et al.~\cite{Niu_2020_nc} used DP to construct the phase diagram of liquid Ga, $\alpha$-Ga, $\beta$-Ga, and Ga-II, in good agreement with experimental results.
In addition, the local structure of liquid Ga and the nucleation process into $\alpha$-Ga, $\beta$-Ga were also studied~\cite{Niu_2020_nc}. 
For carbon, a DP was developed by Wang et.~al.~\cite{Wang_2022_carbon} to simulate the structural properties of 12 different bulk and low-dimensional carbon structures.
For Si, which has both covalent and metallic bonding behavior, the first DP was trained based on datasets generated by classical metadynamics simulations~\cite{Bonati_2018_prl}. 
The DP was then applied to study the crystallisation and the free energy surface between liquid and solid. Many thermodynamics properties near the critical point were found to be close to experimental data. 
Li, et al.~\cite{Li_2021_mtp} trained a DP on DFT data of silicon in the crystalline, liquid, and amorphous phases and thermal conductivity was accurately reproduced. 
DPs were used to study a liquid-liquid phase transition in P, where DP established the main features of the liquid phase diagram~\cite{Yang_2021_prl}. 
In particular, DP phase diagrams of Ga and liquid P clearly indicate how DPs  accurately describe the PES of different phases and transitions between them.   

The applications in elemental bulk systems are among the first applications of DPs in materials science. 
Their successes for different crystal structures and various classes of properties (mechanical, catalytic, irradiation properties, phase transformation, thermal conductivity, et al.) encouraged their extension to multi-element bulk systems and  increasingly complex phenomena.

\subsection{\label{sec:sec3b}Multi-element Bulk Systems}
Al-Mg was the first alloy system for which an accurate DP was developed~\cite{Zhang_2019_prm}. 
This DP was used to describe the 28 crystalline Al-Mg alloys structures in the Materials Project (MP) database~\cite{Jain_2013_aplm}; include accurate prediction of formation energies, equilibrium volumes, elastic constants, vacancy and interstitial formation energies, and unrelaxed surface  energies. 
Wang~\cite{Wang_2020_fc} applied this Al-Mg DP and validated its reliability for crystal structure prediction by using DP+CALYPSO. 
This work   interfaces the DeePMD-kit package and crystal structure prediction software (e.g.,  CALYPSO~\cite{Wang_2012_cpc}, USPEX~\cite{Glass_2006_cpc}, and Pychemia~\cite{PyChemia}). 
Andolina, et al.~\cite{Andolina_2021_prm}  developed a Al-Mg DP based on the original DP to investigate  anisotropic surface segregation. 
Based on the Al-Mg DP, a Al-Cu-Mg ternary DP was developed for the entire compositional space~\cite{Jiang_2021_cpb}. 
2.73 billion alloy configurations were explored in the DP-GEN process.
The resulting DP yields more accurate results for  energetic, mechanical, and defect properties of 58 crystalline structures as compared with  MEAM potentials~\cite{Baskes_1992_prb}. 
The multi-component DP approach can readily be applied to high-entropy alloys for which adequate empirical potentials are difficult to obtain.

Bourgeois, et al.~\cite{Bourgeois_2020_nc} built an Al-Cu DP to simulate the aggregation of vacancies around embedded $\theta'$ precipitates and investigated the nucleation of this strengthening phase onto a template structure. 
An Al-Mg-Zn DP was developed~\cite{Cheng_2020_scripta} and applied to confirm the co-segregation of Mg and Zn atoms at a precipitate and matrix interface. 
DPs have also been found to  be powerful and promising for the prediction of the structure and dynamics of metallic liquids,  glasses, and quasi-crystal~\cite{Wen_2019_prb,Wang_2021_jap,Guo_2019_jac,Guo_2019_njp,Tang_2020_pccp,Tang_2021_acta,Han_2020_prl,Tang_2021_jcp}. 
The DP developed for Pd-Si accurately represented the structure of   liquid and crystal structures, melting points, and glass-forming ability at  compositions near Pd$_3$Si and Pd$_9$Si$_2$ (more accurately than existing EAM potential)~\cite{Wen_2019_prb}. 
Accurate liquid structure and dynamic properties were also obtained with DPs for Nb$_5$Si$_3$~\cite{Wang_2021_jap}, Zr$_{77}$Rh$_{23}$~\cite{Guo_2019_jac}, and Bi$_2$Te$_3$ systems~\cite{Guo_2019_njp}. 

Tang, et al.~\cite{Tang_2020_pccp,Tang_2021_acta,Han_2020_prl,Tang_2021_jcp} performed DP MD simulations of a series Al-based alloys; we focus now on Al-Cr quasicrystals~\cite{Han_2020_prl} as an example and a demonstration of how DPs can be used together with experimental studies. 
Dendritic growth of metastable quasicrystals were observed in the Al$_{13}$Cr$_2$ approximant phase (formed from Al$_{90}$Cr$_{10}$ thin film) by pulsed laser deposition~\cite{Han_2020_prl} which is structurally similar to quasicrystal of the Al$_{13}$Cr$_2$ matrix. 
The Al-Cr DP  was used to simulate the quenching of the  Al$_{90}$Cr$_{10}$ alloy from 2200 to 700 K at  10$^{11}$ K/s. 
There are three types of 13-atom icosahedra in the approximant Al$_{13}$Cr$_2$ phase and one icosahedral  Al-Cr quasicrystal motif. 
These 4 icosahedral motifs are similar despite slightly different Cr-Al bond lengths. 
All 4 types of 13-atom icosahedral motifs were  Cr-centred.  
The icosahedral motif  appears in both the quasicrystal and approximant structures, which results in the survival of the 13-atom icosahedron after laser irradiation. 
This was observed in both simulation and experiment; the success of the DP was attributed to the excellent reproduction of the liquid structure in the DP-based simulations. 

A DP was developed to simulate liquid and glassy silica which proved to have satisfactory accuracy based upon a relatively small training dataset~\cite{Balyakin_2021_pre}. 
Other DPs were developed to calculate transport properties of silicate in the mantle~\cite{Deng_2021_grl,Luo_2021_grl,Luo_2021_gca}. 
DPs were also employed in large-scale calculations of thermodynamic, transport, and structural properties in different molten salts~\cite{Rodriguez_2021_acsami,Liang_2021_jmst,Liang_2020_ats,Pan_2020_cms,Pan_2021_cms,Liang_2021_acsami,Bu_2021_semsc,Zhao_2021_ionics,An_2021_jpcb}. 

Another exciting class of DP applications was for  Li and/or Na-based battery materials~\cite{Xu_2020_jpcc,Huang_2021_jcp,Marcolongo_2019_arxiv,Gupta_2021_ees,Li_2021_icf,Lin_2021_angew};  here we focus on the example of Li$_{10}$GeP$_2$S$_{12}$-type superionic conductors~\cite{Huang_2021_jcp}. 
The DP-GEN scheme was used to generate DPs for three solid-state electrolyte materials (Li$_{10}$GeP$_2$S$_{12}$, Li$_{10}$SiP$_2$S$_{12}$, and Li$_{10}$SnP$_2$S$_{12}$) and applied to  diffusion over a wide temperature range with $\sim$1000 atoms.
The predicted diffusion coefficients slightly overestimated the experimental values but were within the experimental uncertainty.
These DP-based simulations provided a starting point for large size scale and long time scale MD investigations of solid-state electrolyte materials. 

Additional DPs were developed for a wide-range of  other multi-element bulk  systems, including metal oxide~\cite{Andrade_2020_prm,Li_2021_apl,Wu_2021_prb}, metal sulfide~\cite{Balyakin_2022_cms,Wang_2020_arxiv}, thermoelectric SnSe materials~\cite{Guo_2021_mte}, metal borides~\cite{Dai_2020_jecs,Dai_2021_jmst}, and metal carbide~\cite{Dai_2020_jmst} systems. 

\subsection{\label{sec:sec3c}Aqueous Systems}

Since the original DP paper~\cite{Zhang_2018_prl}, water has been widely studied with DPs. 
Ko, et al.~\cite{Ko_2019_mp} applied DPs to perform extensive sampling of  thermal and nuclear quantum fluctuations on an accurate PES. 
In particular, a DP was used to investigate isotopic effects on  structural properties of liquid water (H$_2$O and D$_2$O). 
To understand the experimentally observed isotope effect in the x-ray absorption spectra between liquid H$_2$O and D$_2$O, DP-based, path-integral MD simulations were performed~\cite{Zhang_2020_prb}.  
A more comprehensive study by the same group~\cite{Xu_2020_prb} examined isotope effects on radial distribution functions, O-O-O triplet angular distributions, density and found that the DP-based simulations  were consistent with the experimental observations. 
Recently, Calio, et al.~\cite{Calio_2021_jacs} performed DP-based simulations to interpret experimental observations on the hydrated excess proton in water.

Sommers, et al.~\cite{Sommers_2020_pccp} trained a DP to predict the polarisability of liquid water with \textit{ab initio} accuracy in order to calculate the Raman spectra in long time scale. 
Gartner, et al.~\cite{Gartner_2020_pnas} trained a DP for water to examine the questions of the existence of a liquid-liquid transition in water. 
Andreani, et al.~\cite{Andreani_2020_jpcl} combined neutron scattering experiments and DP MD to investigate hydrogen dynamics in supercritical water. 
The vibrational densities of states observed in DP MD showed coupling between intramolecular vibrational and intermolecular librational and rotational motions. 
Piaggi, et al.~\cite{Piaggi_2021_jctc} used a DP to investigate  ice nucleation in water, hexagonal ice, and cubic ice and obtained quantitative agreement between DP and experiment  (better than the state-of-the-art semi-empirical potentials). 
A more complete description of the phase equilibrium between different phases of H$_2$O was achieved through constructing the DP phase diagram from low temperature and pressure to $\sim$2400 K and $\sim$50 GPa~\cite{Zhang_2021_prl}. 
This is a milestone for DP considering the importance of water, the vast range of temperatures and pressures, and the high accuracy required for free energy representation. 

Tisi, et al.~\cite{Tisi_2021_arxiv} calculated the thermal conductivity of water using both DFT (SCAN) and DP. 
Interestingly, both approaches yield the same conductivity which was  $\sim$50\% higher than  the experimental value.
Zhang, et al.~\cite{Zhang_2021_jpcb} applied DP to help improve the exchange functional in DFT from SCAN to SCAN0 for water. 
Similarly, Torres, et al.~\cite{Torres_2021_jpcb} evaluated the errors of DFT-based simulations on structural and dynamical properties due to time- and size-scale limitations by using DP MD.
In these two examples, the high efficiency and accuracy of DP provided rapid screening of different properties to feedback into DFT exchange-correlation functional optimisation. 
Shi, et al.~\cite{Shi_2021_jpcl} extended DP to produce accurate molecular multipole moments in the bulk and near interfaces consistent with AIMD simulations. 
These moments were used to compute the electrostatic potential at the centre of a molecular-sized hydrophobic cavity in water.

The DP approach has also been used in a wide range of aqueous systems. 
Xu, et al.~\cite{Xu_2019_jpca} developed a DP to perform MD study of zinc ions in liquid water. 
The experimentally observed zinc-water radial distribution function, as well as the X-ray absorption near edge structure spectrum, was well-reproduced by the DP MD simulation. 
Recently, the Limmer group~\cite{Samuel_2021_jcp,Galib_2021_science} applied DP to study  liquid-vapour interfaces. 
They found that the DP yielded accurate  interfacial properties  by incorporating explicit models of the slowly varying long-ranged interactions and training neural networks only on the short-ranged components~\cite{Samuel_2021_jcp}. 
In addition, they trained a DP for solvated N$_2$O$_5$ and bulk ambient water and applied DP MD and importance sampling to study the uptake of N$_2$O$_5$ into an aqueous aerosol~\cite{Galib_2021_science}. 
In contrast to the previous understanding that the uptake process occurs within the bulk of an aerosol, interfacial processes dominate the uptake process due to facile hydrolysis at the liquid-vapor interface and competitive re-evaporation.
This work not only brings new insights to a long-standing questions, but also extends the application of DPs to the liquid-vapor interface. 
Other examples of DP applications in aqueous systems include TiO$_2$-water interfaces~\cite{Andrade_2020_cs}, the ice Ih/XI transition~\cite{Piaggi_2021_mp}, and dynamical states of high-pressure ice VII~\cite{Ye_2021_prl}.

\subsection{\label{sec:sec3d}Other Systems}
Here we briefly list several applications of DPs to other classes of systems including molecular systems, clusters, surfaces, and low-dimensional systems. 
Jiang, et al.~\cite{Jiang_2021_arxiv} developed DPs for sulfuric acid-sulfuric acid, dimethylamine-dimethylamine, and sulfuric aciddimethylamine organic molecular systems to investigate the atmospheric aerosol nucleation process. 
Zeng, et al.~\cite{Zeng_2021_ef} trained a DP based on a dataset for the pyrolysis of $n$-dodecane and performed a reactive DP MD simulation to reveal the detailed pyrolysis mechanism, in good agreement with  experiment. 
Chen, et al.~\cite{Chen_2018_jpcl} used a DP to accurately represent the ground- and excited-state PES of CN$_2$NH. 
This DP  accurately reproduced excited-state topological structures, photo-isomerisation paths, and conical band structure intersections. 
Yang, et al.~\cite{Yang_2021_ct} first generated training datasets through active learning with enhanced sampling and then developed a DP to study the urea decomposition process in water.
Wang, et al.~\cite{Wang_2020_sm} presented a data-driven coarse-grained simulation of polymers in solution and validated the accuracy of this  method with  DPs to construct a coarse-grained potential. 
Pan, et al.~\cite{Pan_2021_jctc} extended the DP-approach to incorporate  external electrostatic potentials in a molecular system; the resultant DP was accurate for energies and forces of representative configurations along the Menshutkin and chorismate mutase reactions pathways. 

A study of metal and alloy clusters and surfaces demonstrated a conflict between Al bulk and cluster energies~\cite{Tuo_2020_jcp}.
This indicates that the compromise between properties are, on occasion, necessary and suggest that DPs should be developed for target properties (this is the specialisation discussed above). 
Andolina, et al. applied a DP to study the nucleation and growth of  seeded core-shell Ag and Au clusters~\cite{Andolina_2021_jpcc} and predict the anisotropic surface segregation for Al-Mg alloys~\cite{Andolina_2021_prm}. 
Wang, et al.~\cite{Wang_2021_msmse} successfully applied DPs in Ag-Au, Au \{111\} surface reconstruction and segregation of Au on the Ag-Au nanoalloy surfaces.

Achar, et al.~\cite{Achar_2021_jpcc} proposed a DP for graphane and showed that it outperforms empirical interatomic potentials for phonon density of states, thermodynamic properties, velocity autocorrelation function, and stress-strain curve up to the yield point.
Wu, et al.~\cite{Wu_2021_prb2} developed a DP for  In$_2$Se$_3$ monolayers and used it to predict a range of thermodynamic properties of In$_2$Se$_3$ polymorphs and lattice dynamics of ferroelectric In$_2$Se$_3$ with \textit{ab initio} accuracy.
Chen, et al.~\cite{Chen_2021_an} applied a DP to simulate the synthesis of amorphous CoFeB during a rapid cooling process. 
The applications of DPs in low-dimensional systems are in their early stage, but existing evidence suggests that  the DP method is promising for the simulation of  low-dimensional materials.

\section{\label{sec:sec4}Accuracy and Efficiency of DPs \\ In Practice}
The summary of recent applications of the DP method in different systems suggests the wide-applicability and high accuracy of DPs.
Nonetheless, it is appropriate to return to the competition between DP accuracy vs the efficiency of DP-based simulations.
On the accuracy issue, we reflect on what we have learned from the applications performed to-date and our experience, to address: (i) How good is DP? (ii) What have we learned? (iii) When can we rely on DPs and when can we not? 
On the efficiency issue, we summarise how fast DPs are by comparison with other approaches from pair potentials to DFT.


\subsection{\label{sec:sec4a}Accuracy}
\subsubsection{\label{sec:sec4a1}How good are DPs?}
Broadly speaking, DPs are more accurate than other types of empirical interatomic potentials;  this was the main reason behind the development of the DP method and its application to a wide range of systems. 
The improvement in accuracy often leads to ``qualitatively" new results. 
We return to  two examples where DPs are  ``qualitatively" better. 
The first example is metal Ti~\cite{Wen_2021_npj}. 
Previous experiments and DFT calculations~\cite{Clouet_2015_nm}  confirmed that the screw $\langle \mathbf{a} \rangle$ dislocation is more stable on the Pyramidal I than on the Prism plane. 
DFT calculations, using pseudopotentials with different numbers of valence electrons~\cite{Poschmann_2017_msmse}, showed that the dislocation core energy of the screw $\langle \mathbf{a} \rangle$ dislocation on the Pyramidal I plane is 18.4 meV/b lower than that on the Prism plane. 
This energy ordering is a prerequisite for the phenomena of screw $\langle \mathbf{a} \rangle$ dislocation ``locking" and ``unlocking" observed in experiment at 150 K~\cite{Clouet_2015_nm}. 
While the screw $\langle \mathbf{a} \rangle$ dislocation is more stable on the Pyramidal I plane, it has  a high energy barrier for glide; hence the screw is ``locked" on this plane. 
However, the screw $\langle \mathbf{a} \rangle$ has a much lower glide barrier on the  higher energy  Prism plane; hence on this plane, the screw ``unlocks" and glides easily. 
Unfortunately,  all extant  empirical Ti potentials predict the incorrect energy ordering;  i.e., the  $\langle \mathbf{a} \rangle$ screw is more stable on the Prism plane (including the generally excellent MEAM empirical potential~\cite{Hennig_2008_prb}), in  ``qualitative" disagreement with experiment. 
The powerful representability and flexibility of the DP method allows the Ti DP to reproduce this important feature of defects in Ti and enables new insights through large-scale MD simulations. 

Another example of ``qualitative'' improvements made possible by a DP is the calculation of the phase diagram of water~\cite{Zhang_2021_prl}. 
TIP4P/2005 is one of the most accurate empirical water models available today for phase diagram prediction~\cite{Zhang_2021_prl}. 
At high temperature and pressure, TIP4P/2005 predicts a first-order transition from ice VII to a plastic phase, in which the BCC oxygen sublattice coexists with freely rotating molecules.
This prediction has not been experimentally confirmed. 
The  DP for water predicts a super-ionic ice VII in that region, in agreement  with  recent experimental observations~\cite{queyroux2020melting}.
DP also better reproduces the phase boundaries between ice VI, VII, and VIII, at high pressures better than TIP4P/2005 . 
These are two of many examples of where the improved quantitative predictions of DP enable ``qualitatively" correct phenomena not accessible through other empirical interatomic potentials.

The accuracy of DP is routinely compared with DFT results especially when benchmarks from  widely accepted empirical interatomic potentials do not exist. 
In most examples (and our own experience), the agreement between energy and force (RMSE) values obtained using a  DP and DFT is commonly smaller than 10 meV/atom and 100 meV/\AA, respectively. 
In some systems, such as the water phase diagram, the error is typically  $\sim$1 meV/molecule.
This accuracy should meet the requirements for most applications. 
In particular, almost half of the examples listed in Table~\ref{tab:table3} focus on non-crystalline systems (liquids, amorphous systems; including liquid metallic alloys, and metallic glasses).
DP-based predictions are especially accurate for liquid properties (DFT references) on liquid structures, diffusion coefficients, thermal conductivities, \ldots.

While the flexibility of DPs is desirable for  describing complex potential energy surfaces, the ultimate accuracy of  DPs are often limited by the accuracy of the DFT training set. 
Since DFT is itself an approximation, perfect agreement with nature is not expected.
Improvements are possible with improved exchange and correlation DFT functionals; climbing the Jacob's ladder from LDA to GGA to meta-GGA to hybrid-GGA to fully non-local approaches~\cite{Perdew_2001_acp}.
In practice, the ultimate limit of DP accuracy can often be associated with the DP training sets, the accuracies of which are often associated with the choice of the DFT functionals. This choice is often dictated by the associated computational cost which rises rapidly on climbing the rungs of the Jacob's ladder.

\subsubsection{\label{sec:sec4a2}What have we learned?}
The two examples, above, demonstrate that DPs can lead to very high accuracy results; comparable to the underlying DFT approaches and higher than the vast majority of empirical interatomic potentials. 
The examples, above, also show us that DPs can be employed to add  new understanding and insight in situations which were previously inaccessible to other computational approaches and experimental observation. 
The DP approach is also useful in the development of new \textit{ab initio} methods and pseudopotentials ~\cite{Zhang_2021_jpcb,Torres_2021_jpcb}. 
Finally, the examples showed how DPs can be readily specialised to describe phenomena for which general purpose DPs do not suffice. 

At early stages of DP development and for simple applications (e.g., simulation of  liquid state structure) where the demands on accuracy are not too high, sampling efficiency is not so critical. 
For such simple applications, an initial DP based on a small set of AIMD trajectory training sets, often leads to efficient and accurate results.
However, sampling efficiency is critical when a general-purpose DP is required or  target properties are  subtle (e.g., where energy differences between phases are very small or some defect properties). 
AIMD may not adequately/efficiently sample atomic configurations that represent those associated with properties of interest. 
This may be addressed by starting with the general purposed DP developed through DP-GEN and tweaking it through identification of property-appropriate and incorporation in the  specialisation step, as described above. 
The specialisation process varies with systems and properties of interest and may be viewed as the art of tweaking DPs based upon physical understanding.

\subsubsection{\label{sec:sec4a3}When can we rely on a DP?}
After performing our normal suite of  property testing on a DP (e.g., see~\cite{Wen_2021_npj}), experience shows that such DPs yield reliable results in atomistic simulations - especially compared with DFT calculations that are of insufficient spacial scales or empirical interatomic potentials that are of insufficient accuracy. 
Critical issues for all ML potentials are representability and transferability.
Representability implies the ability of the functional form to accurately reproduce the target properties. 
Transferability is the ability of a potential to describe the properties which were not included in the training process.

The DP approach usually performs well from a representability perspective; DPs are usually able to provide fits that adequately represent all of the training datasets.
In some cases DP failed to distinguish similar configurations (e.g.~the configurations along the transition path of the screw dislocation in BCC W); the representability of the DP can be improved by using more expressive descriptors, such as a three-body embedding descriptor~\cite{Wang_2021_Tungsten_arxiv}.
On the other hand,  transferability can be non-trivial and subtle for DPs. 
Transferability can be classified as \emph{in-distribution} and \emph{out-of-distribution} transferability.
In-distribution or out-of-distribution transferability is the ability of model to interpolate within or extrapolate out of the sampled configuration distribution, respectively.
The in-distribution transferability of DPs trained with the DP-GEN scheme is generally quite good, providing reliable and accurate predictions of configurations similar to those in  the sampled distribution. 
However, ML potentials will fail in the out-of-distribution transferability where the explored configuration is ``far" from those sampled configurations used in training~\cite{Bartok_2018_prx}. 
A simple example of an out-of-distribution issue is a DP trained using only liquid datasets; such a DP  normally shows poor transferability with respect to crystal datasets because there is little overlap between the liquid and solid configuration distributions.
For defect properties, we do not know \emph{a priori} whether defect  configurations represent an in- or out-of the distribution with respect to those sampled by DP-GEN.
Model deviation (Eq.~\eqref{equ31}) serves as a  good indicator of the transferability of a  DP (without the need for additional DFT calculations).
In the case of  transferability failure, the DP can be specialised by adding configurations to the training set that more closely represent the configuration of interest; this is converting out-of-distribution transferability failure to  in-distribution transferability agreement.

In the cases when a user is not confident  whether a DP is transferable, the DP can be used in conjunction with model deviation after validating against relevant DFT or experimental benchmarks.
DPs are, like other empirical potentials subject to the adages that interatomic potentials ``will work only before they fail" or ``will work until they do not''. 
From this perspective, the replacement of DFT by ML potentials including DP are not completely reliable; DFT will remain the method of choice where very high accuracy property prediction is necessary. 
The combination of DFT and DP provides a  practical strategy for the needs of the materials science community, providing the implicit trade-offs between accuracy and computational efficiency.

\subsection{\label{sec:sec4b}Efficiency}
Using the previous DP application examples, we see that (i) on CPUs,  compressed DPs  are  faster than DFT by a factor of over 10$^6$ and  slower than empirical interatomic potentials such as EAM (MEAM) by $\sim$100 (10) times; (ii) on GPUs, DP compression model is slower than  potentials like EAM by,  of order  10 times.
Of course, the actual efficiency is application dependent (especially for comparison with DFT).  
Compressed DPs can be faster than the original DPs by a factor of over 10 and consume an order of magnitude less memory. 
Additional optimisation of the neural networks at the heart of DPs is possible by  optimisation of different operators on the computational graph and through hardware changes~\cite{Lu_2021_arxiv}. Both DPs and empirical potentials show  linear scaling with the number of atoms on both CPU and GPU machines. 
This linear scaling is the enabler of large-scale atomistic simulations. 
The lower speed of DPs, compared with empirical potentials, is reasonable considering the vast number of parameters in DPs (often at the order of 10$^6$).
Because DPs are and will continue to be slower than empirical potentials, simpler, empirical potentials will continue to play an important role in materials science. 
The  ``competition" between MLs and simpler, empirical potentials drives the continuous improvement of each.
The ML potential community continues to focus on  improving the potential efficiency (computational speed), while empirical potential development continues to develop new formalism increasing  accuracy. 
The concept of ``ML potentials guided by the physics in the empirical potentials" is also an exciting area; e.g., see the recent work of  Mishin, et al.~\cite{Pun_2019_nc,Mishin_2021_acta} .

\section{\label{sec:sec5}Conclusions and outlook}
With  increasing need for atomistic simulations with higher accuracy, larger length scales, longer time scales and computational efficiency, machine learning-based interatomic potentials  are rapidly gaining acceptance in the broad materials science community.
This is especially true in areas where the phenomena of interest are subtle and those in which the material system is complex. 
In this review, we examined the Deep Potential (DP) approach (for ML potentials); summarising the basic theory, how to develop DPs and apply DP-development software and database, how to make DPs more efficient in applications, how to specialise DPs for subtle application, reviewed several DP applications, and discussed DP accuracy and efficiency. 
After several years of evolution, the DP method is now relatively mature, yet continuing to improve in both accuracy and efficiency within an open-source community framework. 
We envision the DP method to continue developing  in the coming years and the continued expansion of the data base of useable DPs.

Continued development will likely proceed along several avenues.
The first is the development of new and more intelligent descriptors for better predictability. 
We  see from the   W example~\cite{Wang_2021_Tungsten_arxiv} that the Peierls barrier (the barrier for dislocation glide) can only be accurately reproduced by expanding the DP descriptor to include three-body embedding.
We suspect that such examples will continue to arise as DPs are expanded to include a broader set of applications in different materials. 
Another issue is related to magnetism; how can magnetic moment degrees of freedom be incorporated into DPs? 
Empirical EAM and MEAM potentials deal with this issue through a set of assumptions and approximations.
More intelligent  descriptors also improve the ease of DP training. 
For example, questions arise for the current hybrid descriptors, with two-body and three-body embeddings about how much to weight three-body embeddings in the hybrid descriptors. 
Current strategies in this area tend to be  based largely on empirical experience; hence, there are opportunities to transform this into a machine-driven process.

An important second area of development will be  improvement of the automation of DP  training and specialisation. 
In  DeePMD-kit and DP-GEN software,  different settings do, on occasion,   influence the performance of trained DPs. 
Although we  presented our experience on choosing these settings in Section~\ref{sec:sec2c}, ideally this experience should be replaced in future generations of the  DeePMD-kit and DP-GEN software. 
The automatic selection of the trust levels in  DP-GEN is already in the testing stage. 
The specialisation step should also be more automated to reduce user intervention in determining: (i) what types of specialisation datasets are needed? (ii) how many specialisation datasets are needed to combine with   DP-GEN datasets? (iii) when to include specialisation datasets and how to modify DeePMD-kit and DP-GEN settings? 
More automated training and specialisation schemes would accelerate the development of new DPs for more systems and applications.

A third area of DP future development is further optimisation of the computational speed of DP. 
Currently, many empirical potentials are faster than DP (at least a factor of 10) which leads potential users to prefer empirical potentials when accuracy demands are not high. 
In principle, DPs cannot be faster than empirical potentials considering the vast number of parameters involved, but decreasing the computation efficiency loss in using a DP can change the  speed - accuracy tradeoff and enable more effective material simulations.
As seen above, the greatly improved accuracy of  DPs over most empirical potentials opens the door to applications where empirical potentials are simply  ``qualitatively" incorrect. 

Finally, the  open-source DP Library (a database for DP including training datasets, training schemes, DP, and testing results) should be greatly expanded to include most of the periodic table, Fig.~\ref{fig:fig6} and alloys. 
This will be an on-going effort requiring contributions from the entire user group. 
Another task is to make the DP Library easier to use.
Both of these will be enabled by improvements in the openness of scientific computing community (e.g., appropriate acknowledgment of contributions).

The development of the Deep Potential approach (and related ML potential approaches) represents an important milestone for the field of atomistic simulations of materials that rests on advances in machine learning technology and descriptors of atomic environments. 
DPs routinely provide high (near DFT) accuracy with reasonable computational efficiency, as compared with empirical potentials. 
The accuracy and efficiency of  DPs open the door to ``qualitatively'' new applications of atomistic simulation.


\section{\label{sec:sec6}Acknowledgments}

TW and DJS gratefully acknowledge the support of the Research Grants Council, Hong Kong SAR, through the Collaborative Research Fund  project number 8730054. 
The work of HW is supported by the National Science Foundation of China under Grant No. 11871110 and 12122103. 
The work of WE is supported in part by a gift from iFlytek to Princeton University.
\bibliography{cited_ref}

\newpage
\beginsupplement
\begin{center}
    \textbf{\large Supplemental Materials}
    ~\\
    ~\\
    \textbf{Brief release history and key milestones of DeePMD-kit}
\end{center}

\begin{table}[!htbp]
	\caption{\label{tab:table1}Brief release history of DeePMD-kit and key milestones are in bold.  }
	\begin{ruledtabular}
		\begin{tabular}{cl}
			\textrm{Version (date)}&\textrm{Main updated features}\\
			\colrule
			v0.x  & Non-smooth and smooth descriptor $\mathcal D^{(2,a)}$\\
			(2018/5-2019/10) & Training and C++ inference \\
			& \\
			v1.0.x & Python inference\\
			(2019/10-2019/11) & Multi-GPU support for MD\\
			&Compatible to TensorFlow v2\\
			&\\
			\textbf{v1.1.x } & \textbf{Offline package and conda installation} \\
			(2019/11-2020/6)& Optimised GPU support for descriptors\\
			&\\
			\textbf{v1.2.x } & \textbf{Polarisability and dipole fitting}\\
			 (2020/5-2021/2)& 
			Fast GELU\footnote{Gaussian Error Linear Unit} activation function\\
			 & Conversion tool for different compatibility\footnote{DeePMD-kit guarantees that codes with the same first two digits in version number are compatible. For example, v1.2.0 is compatible to v1.2.4, but not to v1.3.0 or v2.0.0.}\\
			&\\
			\textbf{v1.3.x} &Compiling with Cuda 11.0 and 11.1\\
			(2021/1-2021/3) & Fix LAMMPS and GPU memory issues\\
			&\\
			\textbf{v2.0.0 (2021/8)} & \textbf{Model compression}\\
			&\textbf{Distributed training with multi-GPU}\\
			&\textbf{New descriptor $\mathcal D^{(3)}$ and hybridisation}\\
			&\textbf{Atom type embedding}\footnote{ Atom type is embedded in a feature space.
			One embedding net and one fitting net are shared with atoms with different types, which reduces the training complexity. Especially useful for multi-component systems.}\\
			& Support ROCM devices\\
			v2.0.1 (2021/9) & Initialise from a compressed model \\
			v2.0.2 (2021/9) & Support the plugin for GROMACS\\
			v2.0.3 (2021/10) & Change default LAMMPS version\\
			&to \texttt{stable\_29Sep2021}\\
		\end{tabular}
	\end{ruledtabular}
\end{table}

The brief release history of DeePMD-kit package and main updated features can be found in Table~\ref{tab:table1}. 
For the first versions v0.x, the non-smooth and the smooth descriptor $\mathcal D^{(2,a)}$ were provided. 
Model training and only the C++ model inference interface were available.
The version v1.0.x was released to provide several features including the Python inference, multi-GPU support for the MD in LAMMPS, and the code compatibility to TensorFlow v2. 
In particular, multi-GPU MD extends the use of DP in larger-size and longer-time scale MD simulations.
For versions before v1.1.0, installation can only be made from source code, which needs the compiling and installation of Python and C++ codes, and the installation of dependencies like the TensorFlow C++ library.
Such installation procedure is not friendly to beginners and limits the applications.
From version v1.1.0, DeePMD-kit supports the installation via the offline package and the conda package manager, which makes the installation as simple as a single click.
For version v1.1.x, GPU support for descriptors was also optimised. 
From version v1.2.0, polarisability and dipole fitting, fast GELU activation function (different from the default tanh activation function), and conversion tool for model compatibility were added. 
DeePMD-kit v1.2.0 is an important milestone where the package can be easily installed under different computing environments, shows good stability, and related research started to boost since then. 
Based on v1.2.x, v1.3.x made further improvements and kept updating with the development of software and hardware. 
CUDA 11.x is supported since version v1.3.x and many non-trivial LAMMPS and GPU memory issues were fixed.

Another milestone for DeePMD-kit comes with v2.0.0. 
First, model compression~\cite{Lu_2021_arxiv} was introduced to accelerate the DP inference up to a factor of 10 by tabulating the embedding net and merging kernels in the descriptor calculation.
The details and process of model compression are presented in Sect.~\ref{sec:sec2e} of  the main text. 
Second, distributed training with multi-GPU becomes possible with the help of the Horovod framework and a linear scaling of the training process is observed for at least 8 GPU cards. 
Third, the new descriptor, three-body embedding $\mathcal D^{(3)}$ (Eq.~\eqref{equ28}) was added and different descriptors can be hybridised together to give better performance.
Next, atom type embedding concept was also introduced to avoid the excessively complex training process especially for multi-component systems. 
At last, ROCm devices~\footnote{\url{https://www.amd.com/en/graphics/servers-solutions-rocm}} are supported in versions later than v2.0.0. 
Optimisations based on v2.0.0 continue. v2.0.1 supports initialisation and continuing training from a compressed model. 
v2.0.2 allows the plugin for GROMACS and v2.0.3 keeps update with the LAMMPS version to \lstinline{stable_29Sep2021}. 
We encourage users to always update DeePMD-kit to the latest versions to enjoy the exciting features at first hand. 
To deal with the incompatibility problem, there is also a tool in v2.0.x to transform the DP trained with early packages to the model which can be readily used by v2.0.x.

\end{document}